\documentclass[a4paper,11pt]{article}
\pdfoutput=1 

\usepackage{jheppub} 

\usepackage[T1]{fontenc} 

\title{\boldmath On the Role of the $\nu_\tau$ Appearance in DUNE in Constraining Standard Neutrino Physics and Beyond}

\author[a,b]{A. Ghoshal,}
\author[a]{A. Giarnetti,}
\author[a]{D. Meloni}

\affiliation[a]{Dipartimento di Matematica e Fisica, 
Universit\`a di Roma Tre\\Via della Vasca Navale 84, 00146 Rome, Italy}
\affiliation[b]{INFN,
Laboratori Nazionali di Frascati, \\ C.P. 13, 00044 Frascati, Italy}

\emailAdd{anishghoshal1@gmail.com}
\emailAdd{giarnetti.alessio@gmail.com}
\emailAdd{davide.meloni@uniroma3.it}

\abstract{We consider the $\nu_\mu \to \nu_{\tau}$ appearance channel in the future Deep Underground Neutrino Experiment (DUNE)
which offers a good statistics of the $\nu_{\tau}$ sample.
In order to measure its impact on constraining the oscillation parameters, we consider several assumptions on the efficiency for $\nu_\tau$ 
charged-current signal events (with subsequent $\tau \to e$ decay) and the related backgrounds 
and study the effects of various systematic uncertainties. Two different neutrino fluxes have been considered, namely a CP-violation
optimized flux and a $\nu_\tau$ optimized flux.

Our results show that the addition of the $\nu_\mu \to \nu_{\tau}$ appearance channel 
does not  reduce the current uncertainties on the standard 3-$\nu$ oscillation parameters while it can improve in a significant way the sensitivity to the Non-Standard Interaction parameter $|\epsilon_{\mu\tau}|$ and to the new mixing angle $\theta_{34}$ of a sterile neutrino model of the $3+1$ type. }

\begin{document} 
\maketitle
\flushbottom

\section{Introduction}

Neutrino experiments over the last 20 years have established the phenomenon of neutrino oscillations and now we are in the era of precision measurements in the 
leptonic sector. 
Although experiments detecting neutrinos from many sources were able to constrain with a very good precision the oscillation parameters, there are still some open questions. In particular, future experiments will be focused on measuring the CP violation in the lepton sector and on determining the sign of the atmospheric mass splitting. Another ambiguity is present in the value of the mixing angle $\theta_{23}$, since we still 
do not know in which octant this angle lies.

Beside checking for the Standard Physics, neutrinos can also be used to test Beyond Standard Model (BSM) physics. Among several scenarios, Non-Standard neutrino Interactions (NSI) with matter \cite{Roulet:1991sm,Guzzo:1991hi} and  the existence of a fourth sterile neutrino
have recently attracted a lot of interest,  especially in connection 
to the ability of long baseline neutrino experiments (LBL) to probe for them \cite{Farzan:2017xzy}-\cite{Abe:2019fyx}.

One of the most powerful neutrino experiment that will be built in the near future is the DEEP Underground Neutrino Experiment 
(DUNE)~\cite{Acciarri:2016crz, Acciarri:2015uup}. This experiment consists of a baseline of 1300~km, planned across two
sites in North America;  the near site, situated at the Fermi National Accelerator Laboratory (FNAL),  Batavia in Illinois, will hosts
the Long Baseline Neutrino Facility (LBNF) and the Near Detector (ND). LBNF~\cite{Strait:2016mof} will provide a GeV-scale $\nu_\mu$ beam 
(with contamination of $\nu_e$) at 1.2~MW, later upgradeable to 2.4~MW. At the opposing end of the baseline, the far site in Sanford Underground Research 
Facility (SURF) in South Dakota will house four 10~kt Liquid Argon Time Projection Chambers (LArTPC) as the Far Detector (FD). 


The DUNE neutrino beam will be able to operate in the  Forward Horn Current (FHC, $\nu$ mode
) and Reverse Horn Current (RHC, $\bar \nu$ mode) modes, in order to look for oscillations of both neutrinos and anti-neutrinos.

DUNE has been designed in order to answer all the questions mentioned before. The proposed neutrino flux (to which we will refer throughout the rest of the paper as the {\it the standard flux} 
\cite{Alion:2016uaj}) will be optimized for the CP violation measurement  and for
this reason will provide a relatively large sample of $\nu_e$ coming from $\nu_\mu \to \nu_e$ oscillations.
However this experiment will also be able to collect a huge $\nu_\tau$ sample, even if most of the neutrinos will not reach the threshold energy of 3.4 GeV for the $\tau$ production. 

Motivated by the interest on $\tau$ neutrinos recently triggered by the observation of 8 $\nu_{\tau}$ interactions by the OPERA experiment \footnote{OPERA observed 10 $\nu_\tau$ candidates, with an expected background of 2 events.} \cite{Agafonova:2014bcr, Galati:2018zov, Agafonova:2018auq},
in this paper we consider in detail the effect of adding the  $\nu_{\mu} \rightarrow \nu_{\tau}$ appearance channel to the more widely used $\nu_e$ appearance and $\nu_\mu$ disappearance modes \footnote{An introductory study about the $\nu_\tau$ appearance in DUNE can be found in \cite{Rashed:2016rda}.}
in the study of the sensitivity of the DUNE experiment to the oscillation parameters of the standard 3-$\nu$ framework as well as in the investigation of the parameter space of the NSI and of the
sterile $3+1$ neutrino models. Differently from \cite{deGouvea:2019ozk}, which focused on the $\tau$ hadronic decays, 
we consider the $\tau \to e$  leptonic decay. In our numerical simulations, performed with the help of the GLoBES software \cite{Huber:2004ka,Huber:2007ji},
we take into account various detection efficiencies, signal to background ratios
(S/B) and systematic uncertainties and explore the performances of the DUNE far detector not only for the  standard flux but also for a {\it $\nu_\tau$-optimized flux} as described in \cite{DUNEwebsite} and 
\cite{Bi}.
In both cases we have considered 3.5 years of data taking in the neutrino mode and 3.5 years in the anti-neutrino modes for a total of 7 years.

The paper is organized as follows: in section \ref{simdetail} we describe the neutrino fluxes and the efficiencies, systematics and backgrounds of the $\nu_\tau$ appearance channel; 
in section \ref{stphys} we focus  on the sensitivity reaches of the DUNE detector on the standard oscillation parameters allowed by the $\nu_{\tau}$ channel alone and for both standard and optimized fluxes.  
section \ref{sect:nsi} is devoted to investigate the impact of this additional
channel on NSI parameter sensitivities while in section \ref{sec:sterile}
we study in detail the sterile neutrino case.  We draw our conclusion in section \ref{concl}.

\section{On the simulation of the \texorpdfstring{$\nu_\tau$}{nutau} appearance in DUNE}
\label{simdetail}
In this section we  specify the differences between the two fluxes used in our numerical simulations and comment on the values for efficiencies, systematics and backgrounds of the $\nu_\tau$ appearance channel taken into account while evaluating the DUNE performance on the mixing parameters measurements, computed for 3.5 + 3.5 years of running time.
\subsection{Fluxes}
\begin{figure}[h]
\begin{center}
\includegraphics[height=7.5cm,width=7.5cm]{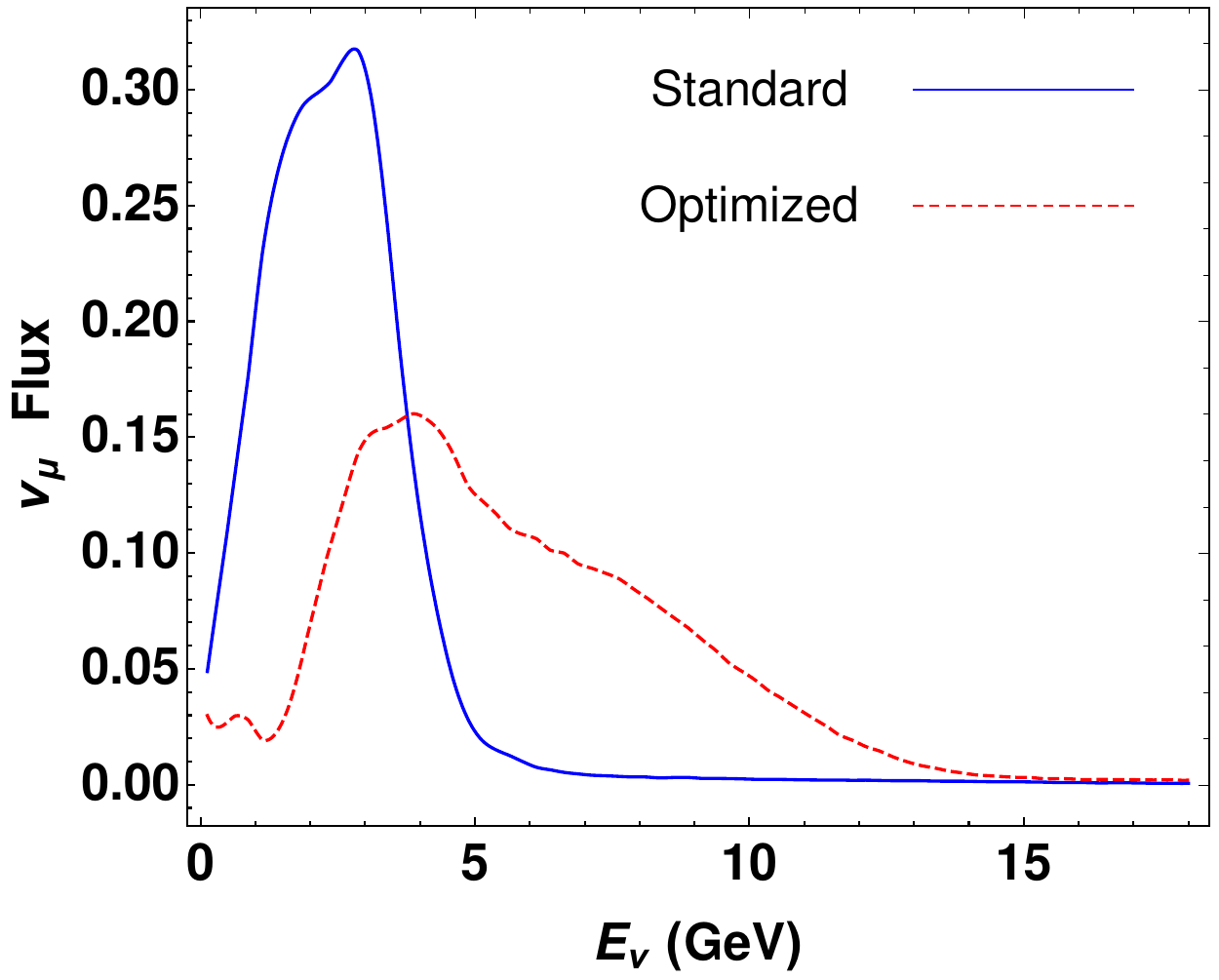} 
\includegraphics[height=7.5cm,width=7.5cm]{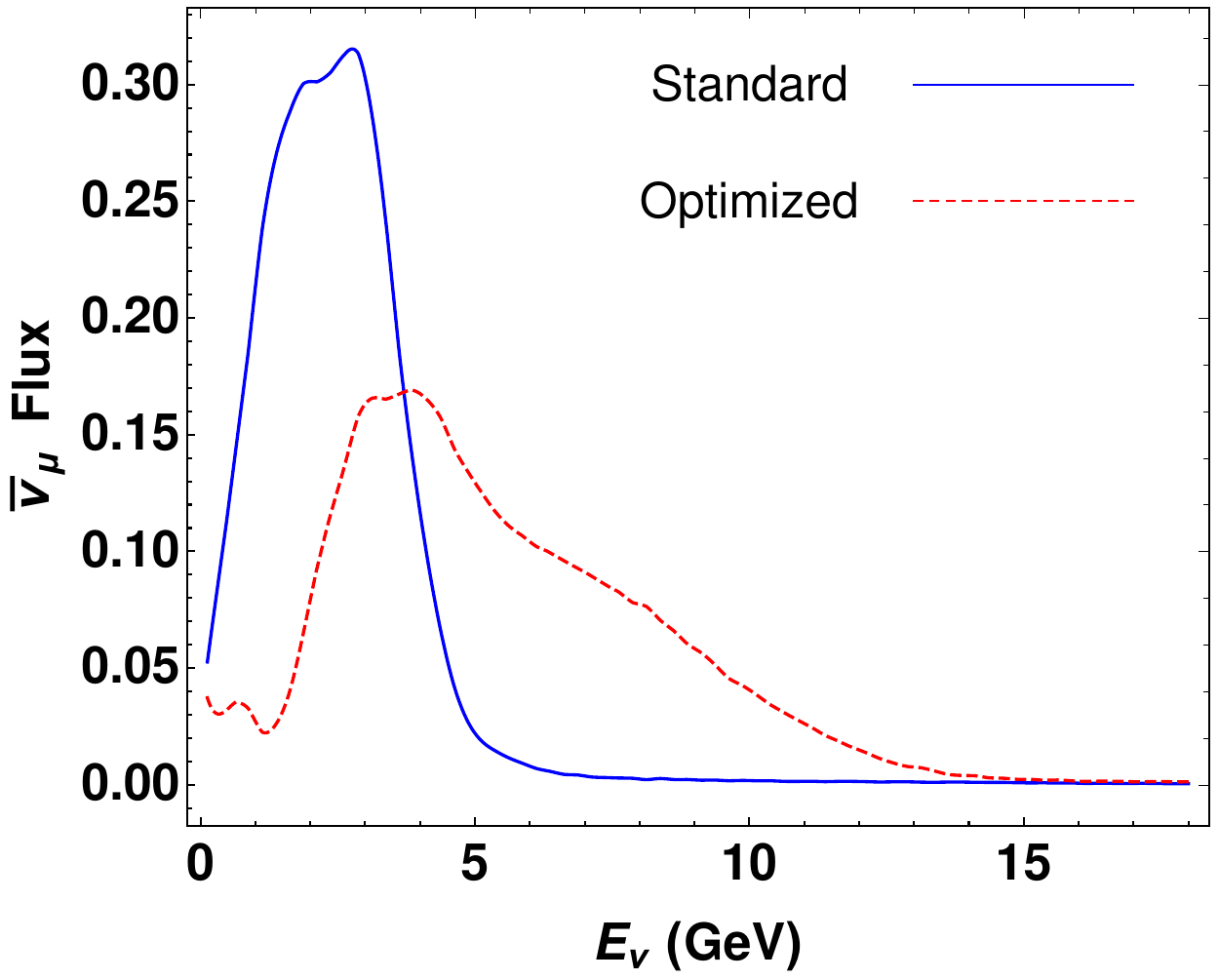} 
\caption{\it  $\nu_\mu$ (left panel) and $\bar{\nu}_\mu$ (right panel) fluxes in  arbitrary units. Standard (blue, solid) and optimized (red, dashed) cases are shown.
} 
\label{fig:flux}
\end{center}
\end{figure}

The two $\nu_\mu$ fluxes discussed in this paper have been displayed in figure \ref{fig:flux}; in both panels (neutrinos on the left, anti-neutrino on the right panel), the blue-solid line refers to the DUNE standard flux \cite{Alion:2016uaj}, which is said to be optimized to maximize sensitivity for CP violation measurements \cite{Acciarri:2015uup}, while the red-dashed case refers to the optimized 
$\nu_\tau$ scenario of \cite{DUNEwebsite} and \cite{Bi}.

The standard flux consists of LBNF beam delivering 1.47 $\times$ 10$^{21}$ protons on target (POT) per year with 80 GeV 
energy running with 1.07 MW beam power and having 1.5 m NuMI (Neutrinos at the Main Injector) style target. The $\tau$ optimized flux is as per proposed by the DUNE collaboration \cite{DUNEwebsite} and it consists of 1.1 $\times$ 10$^{21}$ protons on target (POT) per year with 120
GeV energy running with 1.2 MW beam power and having 1 m NuMI style target. 
The expected un-oscillated charged current (CC) event rates are reported in table \ref{fig:table_flux}; for any of the flux options, we computed the $\nu_e$ and $\nu_\mu$ events (together with their CP conjugate modes) when the LBNF beam is working in the FHC and RHC modes. 
\begin{table}[h]
\begin{center}
\begin{tabular}{c|c|c|c|c|}
\cline{2-5}
                                           & \multicolumn{2}{c|}{\textbf{Standard Flux}} & \multicolumn{2}{c|}{\textbf{Optimized Flux}} \\ \cline{2-5} 
                                           & FHC mode        & RHC mode        & FHC mode         & RHC mode        \\ \hline
\multicolumn{1}{|c|}{$\nu_{\mu}$ CC}       & 30175             & 3225                    & 85523              & 4933                    \\ \hline
\multicolumn{1}{|c|}{$\bar{\nu}_{\mu}$ CC} & 1025              & 9879                    & 1256               & 26221                   \\ \hline
\multicolumn{1}{|c|}{$\nu_{e}$ CC}         & 371               & 136                     & 856                & 258                     \\ \hline
\multicolumn{1}{|c|}{$\bar{\nu}_{e}$ CC}   & 44                & 109                     & 84                 & 215                     \\ \hline
\end{tabular}
\caption{\it  Expected $\nu_e$ and $\nu_\mu$ un-oscillated CC event rates at the DUNE 40-kt far detector for a run-time of total 7 years (3.5 years in neutrino mode and 3.5 years in anti-neutrino mode).
Numbers refer to the two flux options analyzed in this paper.} 
\label{fig:table_flux}
\end{center}
\end{table}
Notice that for the normalization of the tau optimized flux we referred to the number of $\nu_{\tau}$ events as reported in Ref. \cite{Bi}.

We observe that the optimized flux in each of the $\nu$ and $\bar{\nu}$ modes gives a larger number of events. This is mainly due to the more energetic protons involved than in the standard flux which produce higher energy neutrinos and thus implies larger neutrino-nucleus cross sections.  

\subsection{Efficiencies, systematics and backgrounds}
\label{EffSysBkg}
In this section we quote the relevant  efficiencies, systematics and backgrounds of the transition channels included in our numerical simulations.

For the standard flux and the $\nu_e$ appearance and $\nu_\mu$ disappearance channels, we strictly follow Ref.\cite{Acciarri:2015uup}: the  appearance modes in the proposed DUNE experiment
have independent systematic uncertainties of $2\%$ each, while the $\nu_{\mu}$ and $\bar{\nu}_{\mu}$ disappearance
modes have independent systematic uncertainties of $5\%$. The systematic
uncertainties for the backgrounds are $5\%$ for $\nu_e$ and $\nu_\mu$ CC events, $10\%$ for NC events and $20\%$ for $\nu_\tau$ events.
For the standard flux all DUNE collaboration post smearing matrices generated
using the DUNE Fast Monte Carlo (MC) \cite{Acciarri:2015uup} have been used for the $\nu_{e}$ appearance and $\nu_{\mu}$ disappearance channels to set the detection efficiencies.

For the optimized flux studies, since to our knowledge there are no official DUNE MC simulations available, background and signal efficiencies for $\nu_e$ appearance and $\nu_\mu$ disappearance have been set to constant values, 
obtained from averaging the efficiency factors in various energy bins quoted in \cite{Acciarri:2015uup}. For both fluxes, the energy resolution of the detector is described by resolution functions given by the collaboration in Ref. \cite{Acciarri:2015uup}.

As for the $\nu_{\tau}$ appearance detection, for a {\it zeroth order} study
we envisage the DUNE detector to be similar to the ICARUS  one. The $\nu_{\tau}$ appearance sample is composed of $\nu_{\tau}$ CC interactions resulting from
$\nu_{\mu} \rightarrow \nu_{\tau} $ oscillations. Backgrounds to this channel come from the $\nu_{e}$ CC, $\nu_{\mu}$ CC and NC interactions.
Focusing on the leptonic decay of the tau generated by $\nu_{\tau}$ CC interactions, we consider the electron channel only ($18\%$ branching fraction \cite{Tanabashi:2018oca}) since muonic tau decays are affected by large $\nu_{\mu}$ CC background.
From the ICARUS proposal \cite{Aprili:2002wx} it is clear that choosing the right kinematic cuts, the electron channel background can be reduced to only $\nu_{e}$ CC events
coming from two main components, which are the intrinsic $\nu_e$ beam and $\nu_{\mu} \rightarrow \nu_{e}$ oscillation
\footnote{There is negligible intrinsic $\nu_{\tau}$ component in the beam.}. 
Due to the spatial resolution of the DUNE LArTPC detector, very short tau tracks will not be recognized, and for this reason $\nu_{e}$ CC background cannot be avoided.
In this paper, for both standard and optimized neutrino fluxes, the overall $\nu_{\tau}$ and $\bar{\nu}_{\tau}$ appearance  signal efficiency has been set to two distinct values (including the branching fraction), that are $6\%$ \cite{Barger:2001yxa} and, in order to show the full potential of the electron channel, also the maximum reachable efficiency of $18\%$. 
The former case has to be considered as a pessimistic case, being related to the expected selection efficiency of an old-generation detector like ICARUS; in fact, we expect DUNE to have better performances in reconstructing electrons tracks and in distinguishing electrons coming from tau decays. On the other hand, the latter efficiency of 18\% has to be considered as an optimistic assumption, since it corresponds to DUNE able to reconstruct and recognize all produced electrons.

The number of the $\nu_{e}$ CC background has been set to a constant value as to avoid the parameter dependence of the rates that, with such a comparatively small number of $\nu_{\tau}$ events, can overshadow the effect of the signal in the simulations; in particular, such a background has been chosen in order to reproduce, at the best fit parameter values, two S/B ratios discussed for ICARUS: 18.6 \cite{Aprili:2002wx} and 2.45 \cite{Barger:2001yxa}. 
In order to show the impact of $\nu_\tau$ systematic uncertainties on the results,
we study the situation where such an uncertainty for $\nu_{\tau}$ signal events  is fixed to the same value used by the DUNE collaboration when $\nu_{\tau}$ CC are considered as a background to the $\nu_{e}$ appearance and $\nu_{\mu}$
disappearance channels, that is $20\%$, and also the situation where an optimistic $10\%$ is taken into account.

Expected total rates for $\nu_{\tau}$, $\bar{\nu}_{\tau}$ and $\nu_{e}$, $\bar{\nu}_{e}$ CC events for both fluxes are shown in tables \ref{tab:calc1} and 
\ref{tab:calc2}.

\section{The case of Standard Physics}
\label{stphys}

The $\nu_{\mu} \rightarrow \nu_{\tau}$ oscillation probability, among others, was calculated in \cite{Akhmedov:2004ny}. Neglecting terms containing the solar mass difference $\Delta m_{21}^2=m_2^2-m_1^2$ and the small $\sin \theta_{13}$, in vacuum such a probability reads:
\begin{equation}
 P_{\mu\tau} \approx \cos ^4 \theta_{13} \sin ^2 {2 \theta_{23}}  \sin ^2 \left( \frac{\Delta m^2 _{31} L}{ 4E} \right)\,.
 \label{eq:nutau}
\end{equation}
Eq.(\ref{eq:nutau}) shows that the $\nu_\tau$ appearance channel is particularly sensitive to $\theta_{23}$ and to the atmospheric mass-squared splitting $\Delta m_{31}^2=m_3^2-m_1^2$. However, also the other two channels are expected to be sensitive to the same two parameters since, neglecting solar terms, we have:
\begin{equation}
 P_{\mu e} \approx 4 \sin ^2 { \theta_{13}} \cos ^2 { \theta_{13}} \sin ^2  \theta_{23} \sin^2 \left( \frac{\Delta m^2 _{31} L}{ 4 E} \right)\,,
 \label{prob_e}
\end{equation}
and
\begin{equation}
 P_{\mu\mu} \approx 1 - (\sin ^2 {2 \theta_{23}} \cos ^4 \theta_{13} + \sin ^2 {2 \theta_{13}} \sin ^2 2 \theta_{23} ) \sin ^2 \left( \frac{\Delta m^2 _{31} L}{ 4E} \right)\,.
\end{equation}
In DUNE the mean neutrino energy in the standard flux has been chosen in order to maximize the atmospheric term; since the minimum $\nu_\tau$ energy needed to be converted in a $\tau$ lepton is around 3.4 GeV,  the number of $\nu_e$ and $\nu_\mu$ events will be much bigger than the number of $\nu_\tau$ CC.  For this reason, we expect that constraints on $\theta_{23}$ and $\Delta m_{31}^2$ will be mainly set by $\nu_{\mu} \rightarrow \nu_{e}$ and $\nu_{\mu} \rightarrow \nu_{\mu}$ channels.
Notice also that next terms in the $\Delta m^2_{21}$ and $\theta_{13}$  of eq.(\ref{eq:nutau}) would exhibit a $\sin{\delta_{CP}}$ dependence, so we expect this channel to be also partially sensitive to CP violation searches. However, due to the very large leading term, the changes in probability due to the CP violation phase will be comparatively very small and definitely less important than the corresponding CP violating terms 
in $P_{\mu e}$.

In summary, considering the $\nu_{\mu} \rightarrow \nu_{\tau}$ oscillation probability and the lack of statistics, the $\nu_{\tau}$ appearance channel is expected to have a negligible impact on standard physics studies.

\subsection{Expected rates for signal and backgrounds}
In this section we estimate the event rates for the $\nu_\tau$ appearance in DUNE. We have used mixing parameters with their error bars from 
\cite{Esteban:2018azc} which we summarize in Table 2.
\begin{table}[h]
\centering
\begin{tabular}{|lcc|} \hline 
Parameter &    Central Value & Relative Uncertainty \\ \hline
$\theta_{12}$ & 0.59 & 2.3\% \\ 
$\theta_{23}$ (NH) & 0.866  & 2.0\% \\ 
$\theta_{13}$ & 0.15  & 1.4\% \\ 
$\Delta m^2_{21}$ & 7.39$\times10^{-5}$~eV$^2$ & 2.8\% \\ 
$\Delta m^2_{31}$ (NH) & 2.525$\times10^{-3}$~eV$^2$ &  1.3\% \\ \hline
\end{tabular}
\caption{\it \label{tab:oscpar_nufit2} Central values and relative uncertainty of neutrino oscillation parameters from a global fit to neutrino oscillation data \cite{Esteban:2018azc}. 
As in \cite{Acciarri:2016crz}, for non-Gaussian parameter $\theta_{23}$ the relative uncertainty is computed using 1/6 of the 3$\sigma$ allowed range.
Normal mass hierarchy (NH) is assumed. Throughout the analysis presented in this paper, we assumed true values of $\delta_{CP}$ to be $215 ^\circ$ as per Ref. \cite{Esteban:2018azc}. We have used these values as central values for our simulation unless otherwise stated explicitly in the text.}
\end{table}
The expected rates of the $\nu_\tau$ signal and background (Bkg) from the two fluxes considered here are reported in table \ref{tab:calc1} for the standard flux and in table \ref{tab:calc2} for the optimized flux. These tables show the total number of expected $\nu_\tau$ and $\nu_e$ CC events in DUNE without considering any detection strategy. Using the efficiencies and S/B values discussed in the previous section, the number of signal and background events in every configuration considered in this paper can be obtained. In both tables we specify the two sources of electron backgrounds coming from the intrinsic $\nu_e$ component of the beam, [$\nu_e \oplus \bar \nu_e$ CC Background (beam)], and from $\nu_\mu \to \nu_e$ oscillations, [CC Background (oscillation)]. 
\begin{table}[h]
\begin{center}
\begin{tabular}{ |c|c| } 
 \hline
 \textbf{$\nu$ mode}   &   \\ \hline
 $\nu_{\tau}$ Signal & 277  \\ 
 $\bar{\nu}_{\tau}$ Signal & 26  \\ \hline
 Total Signal & 303  \\ \hline
 $\nu_{e} + \bar{\nu}_{e}$ CC Bkg (beam) & 333 + 38 \\
 $\nu_e + \bar{\nu}_{e}$ CC Bkg (oscillation) &    1753 + 12   \\
 \hline
\end{tabular}
\begin{tabular}{ |c|c|c| } 
 \hline
 \textbf{$\bar{\nu}$ mode}     &  \\ \hline
 $\nu_{\tau}$ Signal & 68  \\ 
 $\bar{\nu}_{\tau}$ Signal & 85  \\ \hline
 Total Signal & 153  \\ \hline
 $\nu_{e} + \bar{\nu}_{e}$ CC Bkg (beam) & 117 + 104\\
 $\nu_e + \bar{\nu}_{e}$ CC Bkg (oscillation) &   90 + 188    \\ 
 \hline
\end{tabular}
\end{center}
\caption{\it \label{tab:calc1} Expected total number of events after oscillation at the 40-kt far detector for Signals and Backgrounds (Bkg) obtained using no selection efficiencies hypothesis in the case of the standard flux and for Normal Hierarchy (NH). $\delta_{CP} = 215 ^\circ $ is assumed \cite{Esteban:2018azc}. The events correspond to DUNE running for a total of 7 years (3.5 years in neutrino mode and 3.5 years in anti-neutrino mode). }
\end{table}

\begin{table}[h]
\begin{center}
\begin{tabular}{ |c|c| } 
 \hline
 \textbf{$\nu$ mode}  &   \\ \hline
 $\nu_{\tau}$ Signal & 2673  \\ 
 $\bar{\nu}_{\tau}$ Signal & 34  \\ \hline
 Total Signal & 2707  \\ \hline
 $\nu_{e} + \bar{\nu}_{e}$ CC Bkg (beam) & 688 + 63 \\
 $\nu_{e} + \bar{\nu}_{e}$ CC Bkg (oscillation) &    1958 + 11   \\
 \hline
\end{tabular}
\begin{tabular}{ |c|c|c| } 
 \hline
 \textbf{$\bar{\nu}$ mode}     &  \\ \hline
 $\nu_{\tau}$ Signal & 98  \\ 
 $\bar{\nu}_{\tau}$ Signal & 983  \\ \hline
 Total Signal & 1081  \\ \hline
 $\nu_{e} + \bar{\nu}_{e}$ CC Bkg (beam) & 176 + 177\\
 $\nu_{e}$ CC Bkg (oscillation) &   76 + 324     \\ 
 \hline
\end{tabular}
\end{center}
\caption{\it \label{tab:calc2} Same as table \ref{tab:calc2} but for the optimized flux.}
\end{table}
These numbers must be compared with a total of 2043 (2369) $\nu_\mu \to \nu_e \oplus \bar \nu_\mu \to \bar \nu_e$ CC signal events for the standard (optimized) flux and with a total of 14206 (67143) $\nu_\mu \to \nu_\mu \oplus \bar \nu_\mu \to \bar \nu_\mu$ CC signal events. 


We clearly observe that the DUNE experiment is able by itself to provide a $\tau$ sample around 300 events in FHC mode and 150 in RHC mode
because of the generous $\nu_\mu$ flux components above the tau production threshold. On top of that, as per the plan for the optimized flux,
there is a huge gain in statistics by roughly a factor of 10 with respect to standard taus, thereby justifying the possibility to explore scenarios of new physics with taus.

\subsection{Details on the \texorpdfstring{$\chi^2$}{chi2} definition}
The confidence regions involving the sensitivity of the measurement of the oscillation parameters are determined based on the standard pull method 
\cite{Huber:2002mx,Fogli:2002pt,Ankowski:2016jdd} as implemented in GLoBES.
The $\chi^2$ is calculated by the minimizing over the nuisance parameters $\vec \xi$. For
every transition channel $c$ with energy bin $i$ (in the case of DUNE the number of energy bins suggested by the collaboration is 71 \cite{Alion:2016uaj}), a Poissonian $\chi^2$ distribution is used of the form:
\begin{equation}
  \chi_c^2 = \sum_i  2 \bigg( f_{c,i}(\vec{\theta}, \vec{\xi}) - O_{c,i}
                            + O_{c,i} \ln \frac{O_{c,i}}{f_{c,i}(\vec{\theta}, \vec{\xi})} \bigg) \,.
\label{equ:chirule}
\end{equation}
For the $i$-th energy bin for a given channel $c$, and for a set of
oscillation parameters $\vec{\theta}$ and nuisance parameters $\vec{\xi}$,
$f_{c,i}(\vec{\theta}, \vec{\xi})$ is the predicted number of events and  
$O_{c,i}$, is the observed
event rate \textit{i.e.}, the event rate considering assumed true values of the 
oscillation parameters.  Both $f_{c,i}$ and $O_{c,i}$ receive
contributions from different sources $s$, that usually involve signal and background rates given by $R_{c,s,i}(\vec{\theta})$, 
such that
\begin{equation}
  f_{c,i}(\vec{\theta}, \vec{\xi}) = \sum_s \left(1 + a_{c,s}(\vec{\xi}) \right) R_{c,s,i}(\vec{\theta}) \,.
\end{equation}
The auxiliary parameters $a_{c,s}$ have the form $a_{c,s} \equiv \sum_k w_{c,s,k} \, \xi_k \,,
$ in which the coefficients $w_{c,s,k}$ assume the values 1 or 0 corresponding respectively to 
a particular nuisance parameter $\xi_k$ affecting or not affecting the
contribution from the source $s$ to channel $c$.
 
Therefore, the $\chi^2$ is given by in total by:
\[
\chi^2=\min_\xi\left\{\sum_{\,c}\chi^2_{c}
+\left(\frac{\xi_N}{\sigma_N}\right)^2\right\},
\]
where the overall signal normalization is represented respectively by the last term known as the pull term. The different values used
for $\sigma_N$ have been quoted in section \ref{EffSysBkg} (for both signal and background).



\subsection{Numerical results}

\paragraph{Standard Flux}
In this section we exclusively use the standard flux configurations with an exposure of 3.5 + 3.5 years for investigating the sensitivity and correlation 
among the standard physics parameters obtained from the $\nu_{\tau}$ appearance channel. 

\begin{figure}[tbp]
\begin{center}
\includegraphics[height=7.5cm,width=7.5cm]{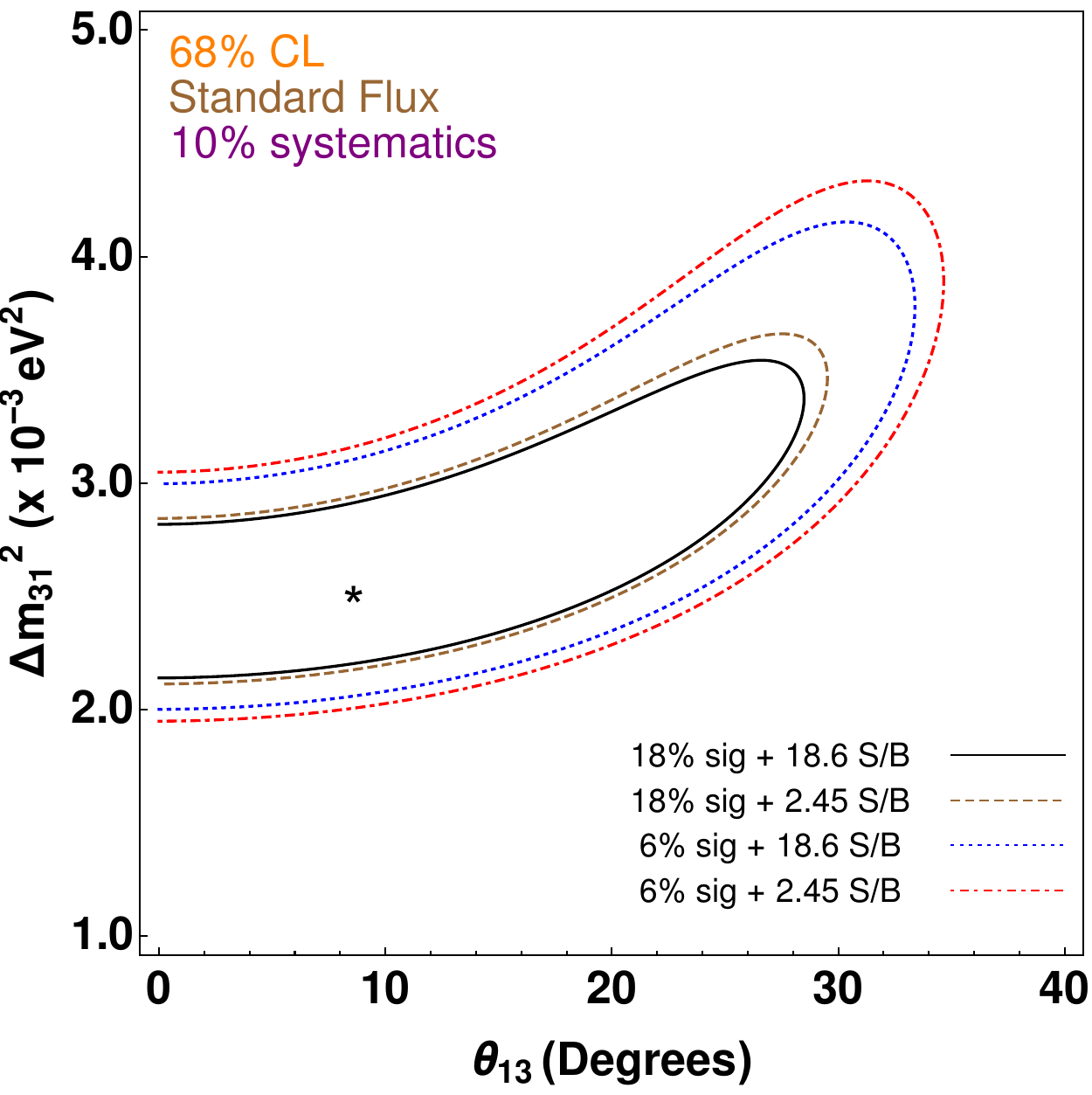} 
\includegraphics[height=7.5cm,width=7.5cm]{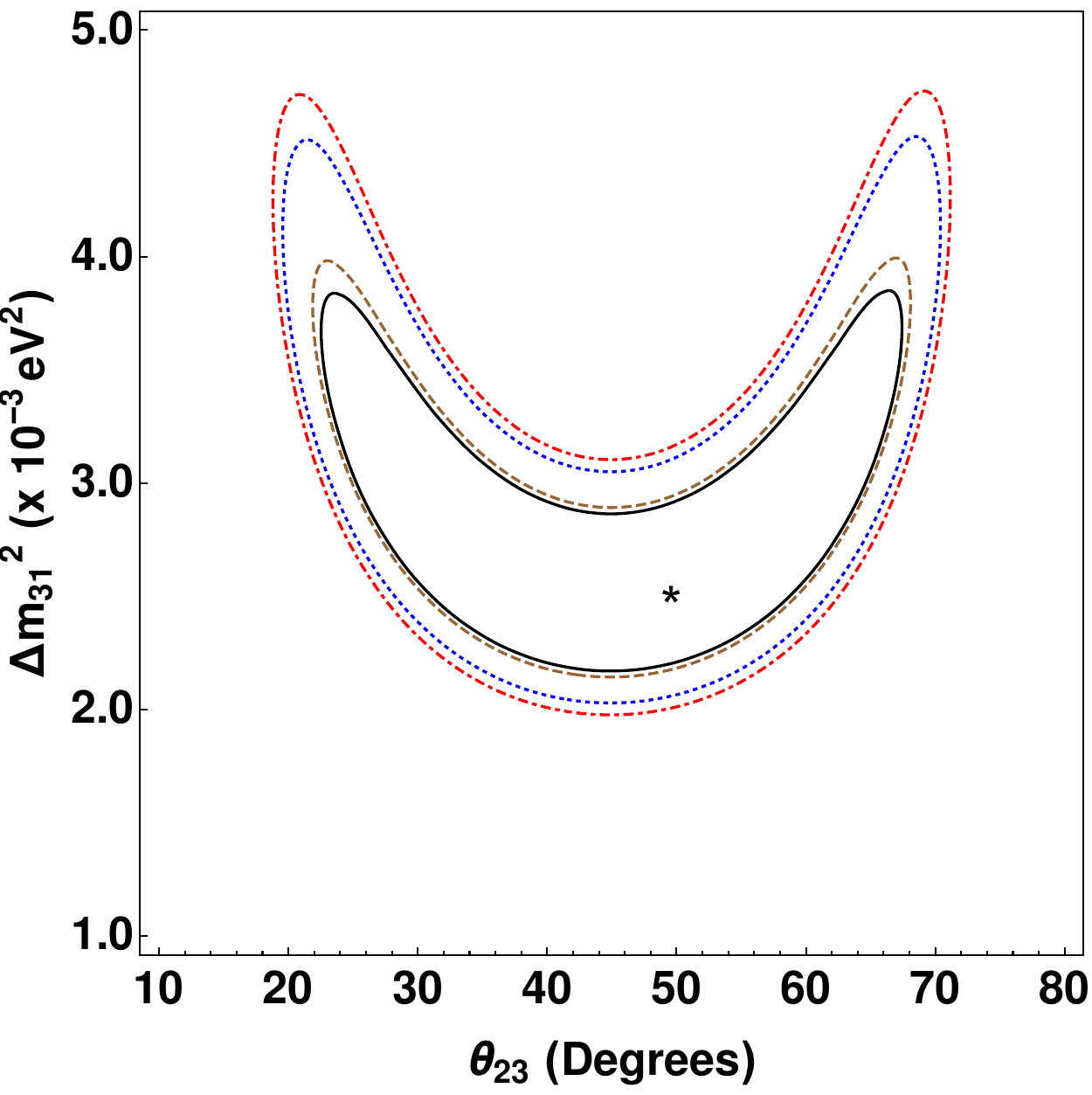} 
\includegraphics[height=7.5cm,width=7.5cm]{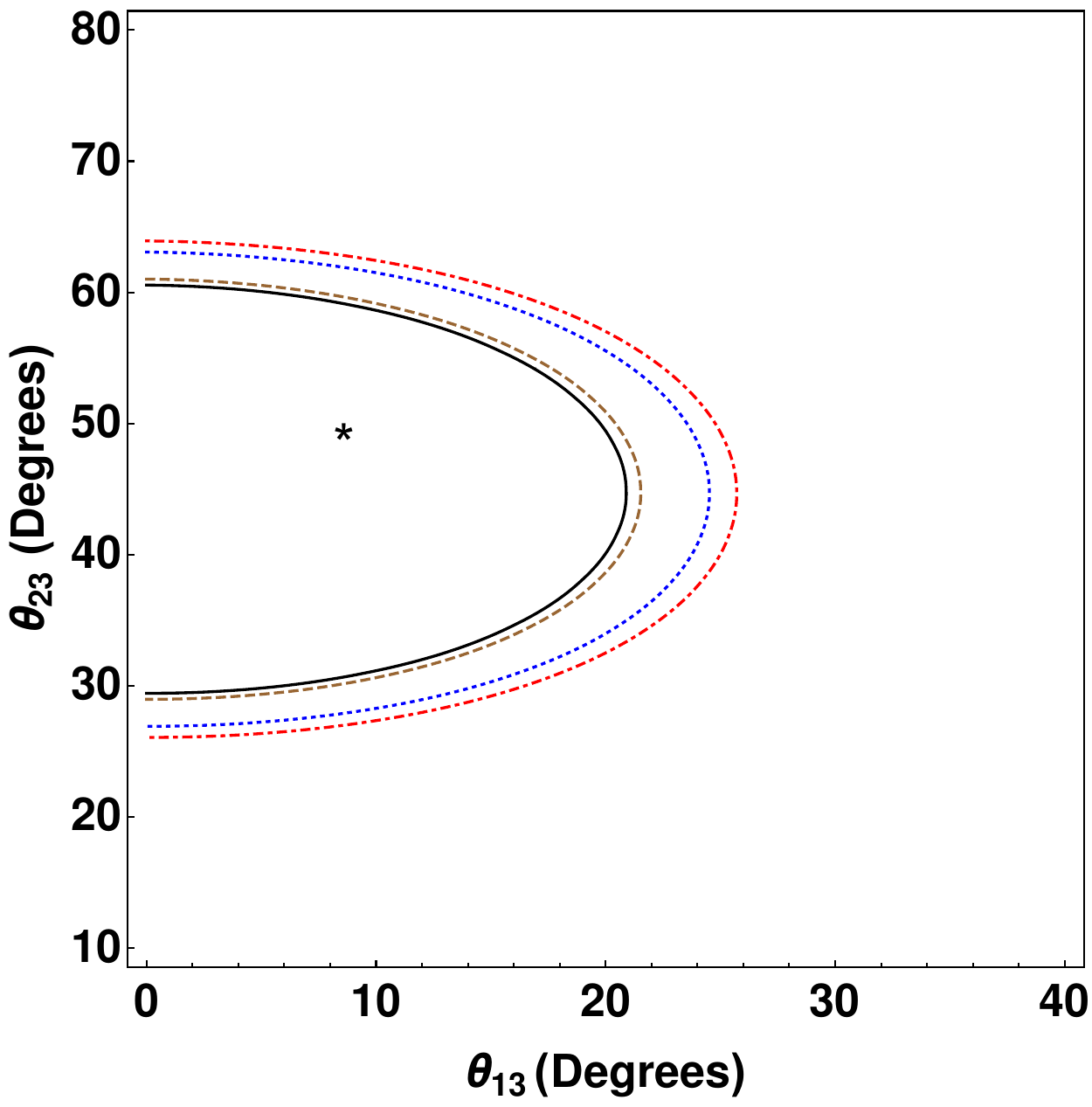} 
\includegraphics[height=7.5cm,width=7.5cm]{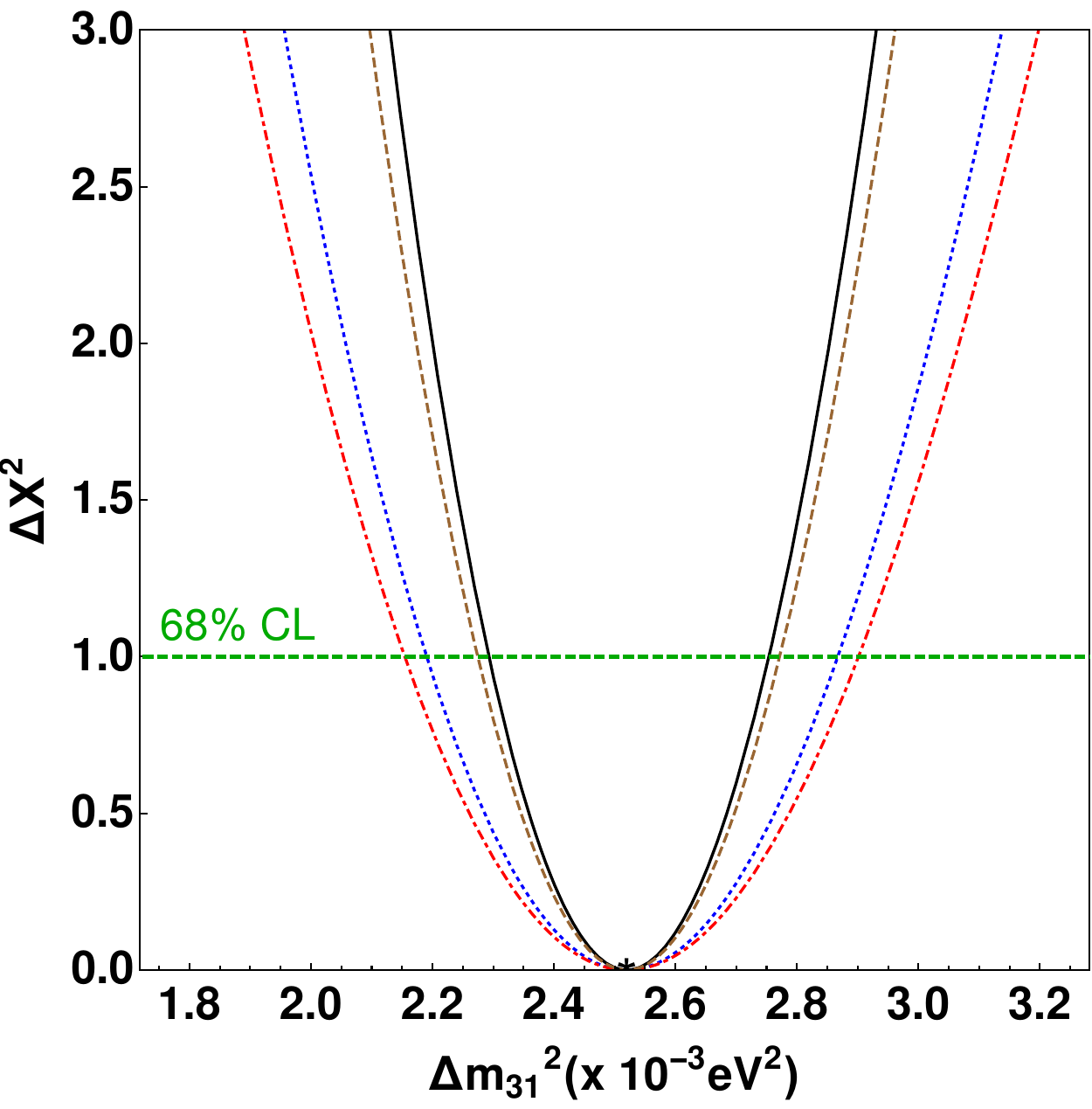} 
\caption{\it  Expected contours at 68\% CL in the oscillation parameter planes (as mentioned in each graph axis) for $S/B = 2.45$, with $\nu_{\tau}$ detection efficiency
of 6\% (Red, DotDashed) and 18\% (Brown, Dashed) and for $S/B = 18.6$ with $\nu_{\tau}$ detection efficiency of 6\% (Blue, Dotted) and 18\% (Black, Solid).
Standard flux has been assumed in the simulation, using only the $\nu_{\tau}$ appearance channel in the Normal Hierarchy case with a  10\% signal uncertainty. 
Marginalization over all undisplayed parameters has been performed.
For the bottom right graph, $\Delta {\chi}^2 = \chi^2 -\chi_{min} ^2$ as a function of the true $\Delta m^2 _{31}$ is plotted.  
Stars represent the simulated true values.
} 
\label{sp_std_10}
\end{center}
\end{figure}

In figure \ref{sp_std_10} we report correlation plots in the planes $(\theta_{13},\Delta m^2_{31})$ (top left), $(\theta_{23},\Delta m^2_{31})$ (top right) and $(\theta_{13},\theta_{23})$ (bottom left), while  the $\Delta \chi^2 = \chi^2 -\chi_{min} ^2$ versus $\Delta m^2_{31}$ is shown in the bottom right panel. In each panel we show four different cases: (Red, DotDashed) refers to $\nu_{\tau}$ detection efficiency
of 6\% and $S/B = 2.45$, (Brown, Dashed) to the same $S/B$ but  18\% of detection efficiencies while the (Blue, Dotted) and the (Black, Solid) lines refer to 6\% and 18\%, respectively, and the same $S/B = 18.6$. Contours are at 68\% confidence level (CL) and are obtained assuming a 10\% systematic error on the signal. 

It is clear that when the $\nu_{\tau}$ detection efficiency is increased from 6\% to 18\%, the number of signal events is increased and this results in smaller allowed regions in the correlation plots. 
The allowed range for $\theta_{13}$ can be reduced by up to 18\%, as we can see in the correlation plot in the $(\theta_{13},\Delta m_{31}^2)$ plane. 
On the other hand, the $\theta_{23}$ range can be reduced approximately by 15\%, as shown in the $(\theta_{23},\Delta m_{31}^2)$ plane. For $\Delta m^2 _{31}$, instead, an improvement of approximately 30\% can be reached passing from the worst case ($S/B = 2.45$, with $\nu_{\tau}$ detection efficiency at 6\%) to the best one ($S/B = 18.6$, with $\nu_{\tau}$ detection efficiency at 18\%). 
Notice that the OPERA experiment measured $\Delta m ^2 _{31}= (2.7 \pm 0.7)\times 10^{-3}$ eV$^2$ \cite{Agafonova:2018auq,Meloni:2019pse} while the DUNE setup here discussed with an 18\% $\tau$ detection efficiency reaches a much better relative uncertainty of about 8\%.

Also a larger $S/B$ gives a better sensitivity to the mixing parameters: in particular, for  $S/B=18.6$ the correlation plots show a reduction of the mixing angles allowed ranges of approximately 5\% if compared to the case $S/B = 2.45$. Roughly the same reduction can be noticed in the relative uncertainty of $\Delta m^2 _{31}$. 
Thus we conclude that an increase in efficiency allows a better performance of the DUNE detector than 
a larger $S/B$.

\begin{figure}[tbp]
\begin{center}
\includegraphics[height=7.5cm,width=7.5cm]{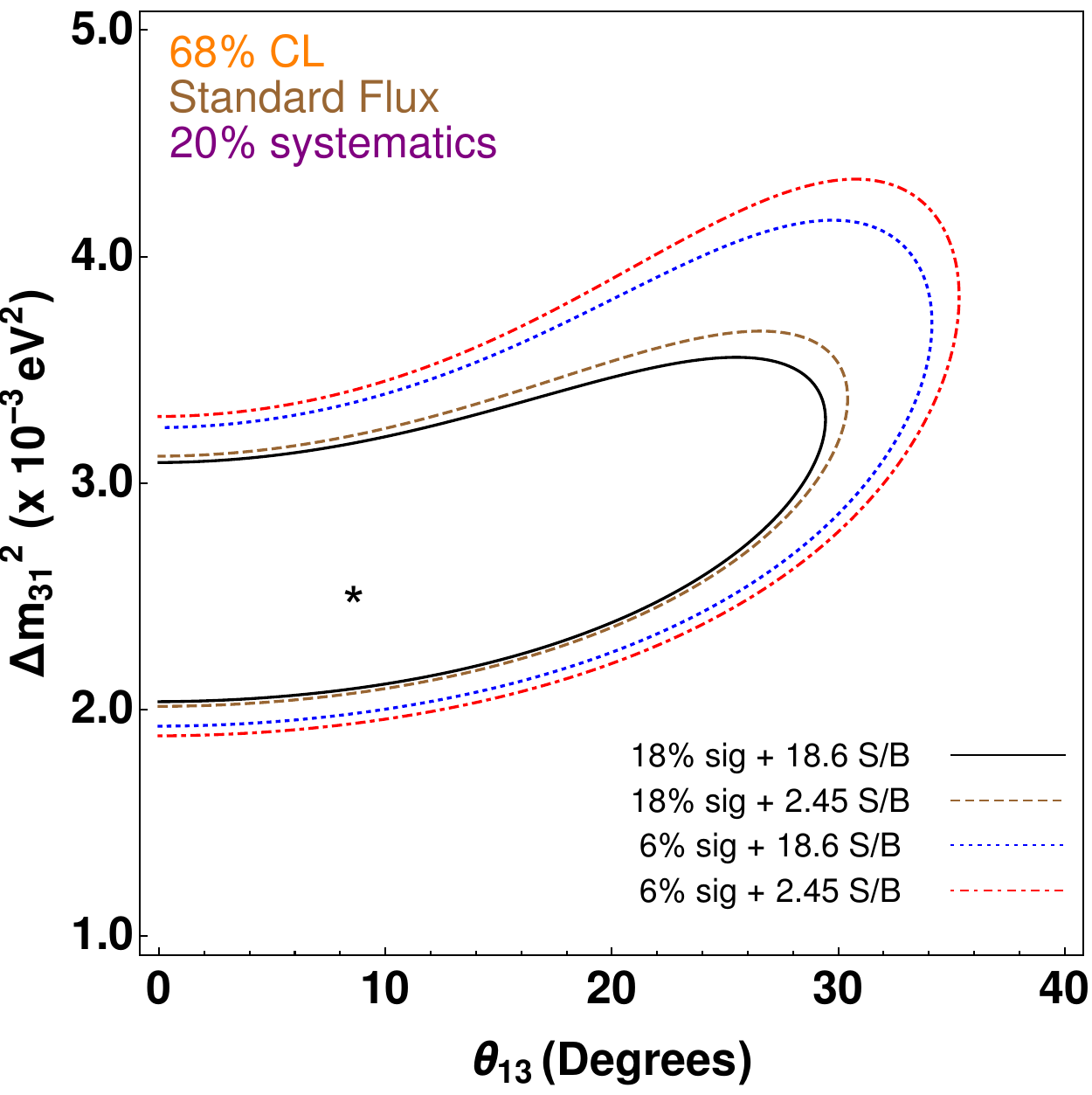} 
\includegraphics[height=7.5cm,width=7.5cm]{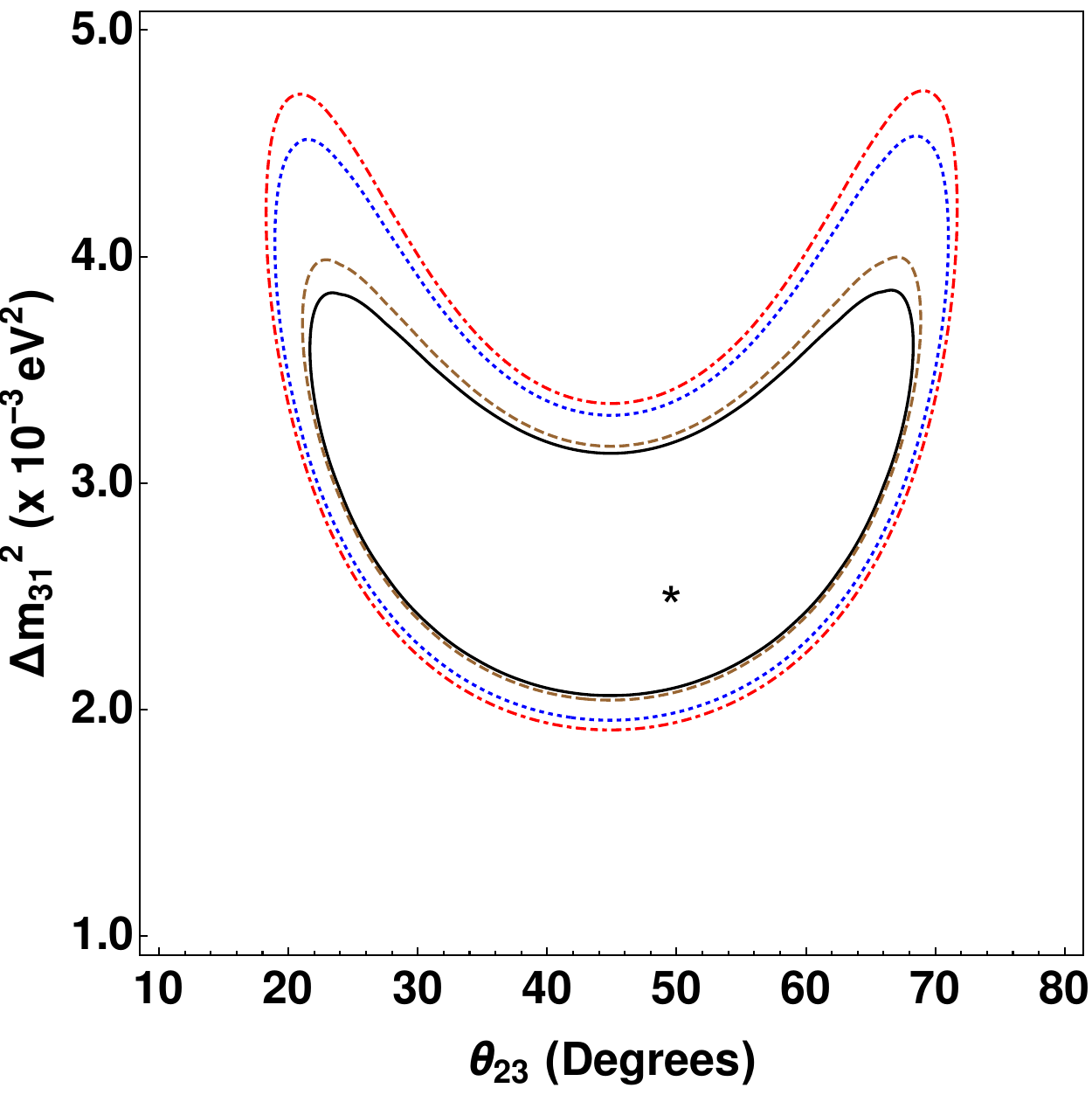} 
\includegraphics[height=7.5cm,width=7.5cm]{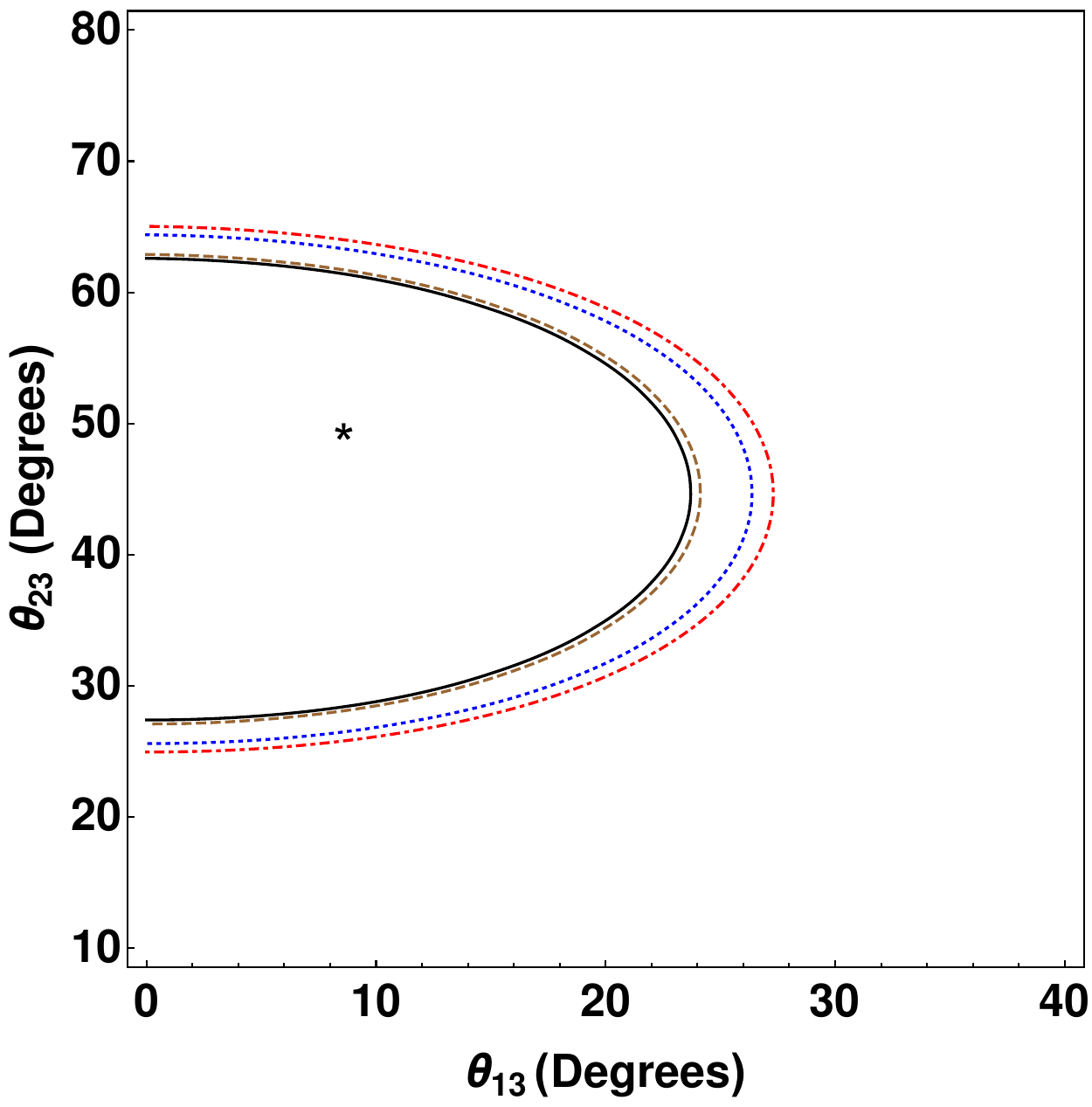} 
\includegraphics[height=7.5cm,width=7.5cm]{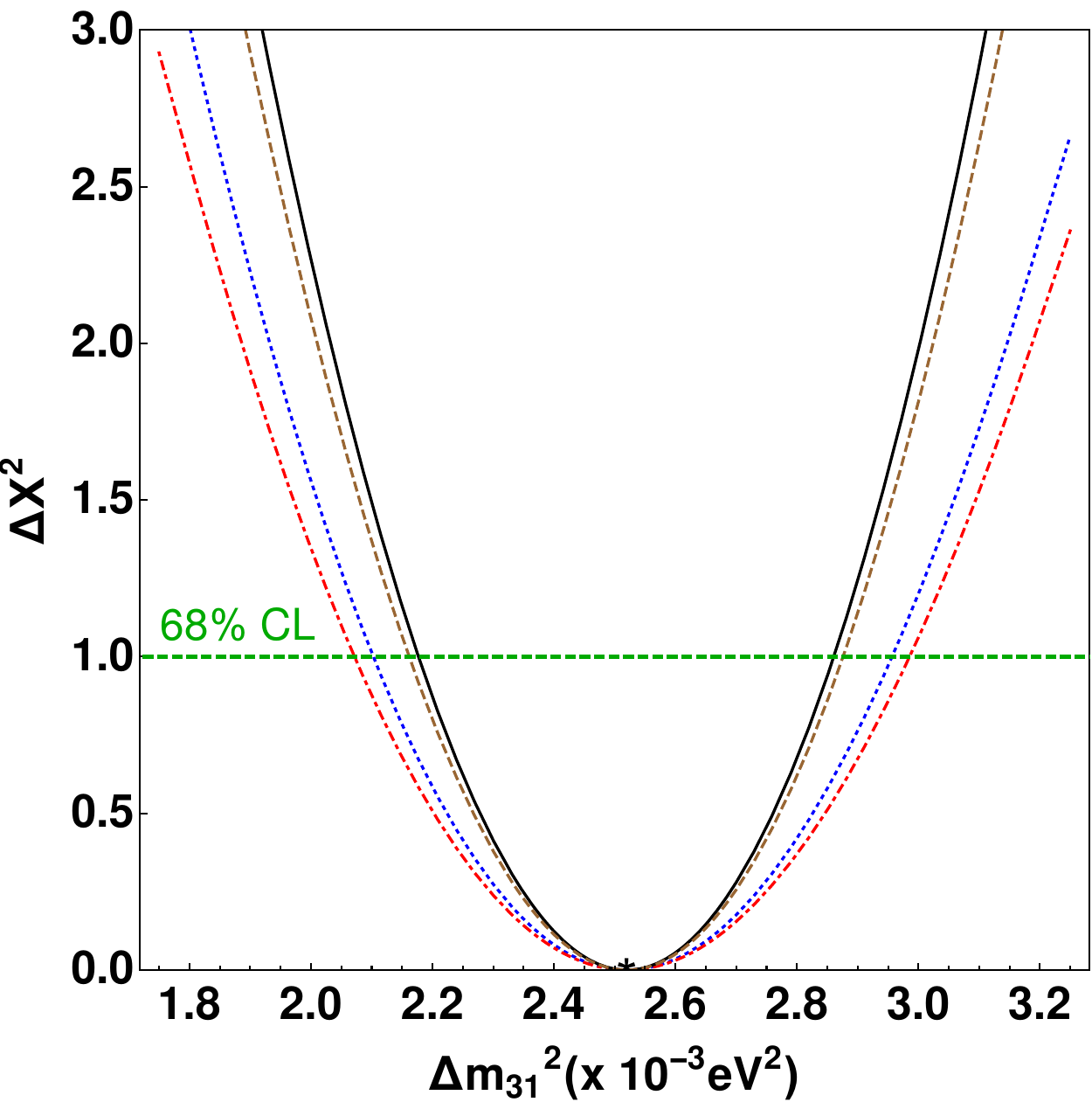} 
\caption{\it  Same as figure \ref{sp_std_10} but with 20\% signal uncertainty.
} 
\label{sp_std_20}
\end{center}
\end{figure}

In figure \ref{sp_std_20} we depict the same plots as in figure \ref{sp_std_10} but with 20\% systematic uncertainty on the signal.
We see that doubling the systematic uncertainty from 10\% to 20\% results in a decrease in sensitivity of approximately 8\% for all mixing parameters.

\paragraph{Optimized Flux}

In this section, we exclusively use the tau optimized flux configurations with an exposure of 3.5 + 3.5 years for investigating the sensitivity and correlation among the 
standard physics parameters as obtained from the $\nu_{\tau}$ appearance channel only.

\begin{figure}[tbp]
\begin{center}
\includegraphics[height=7.5cm,width=7.5cm]{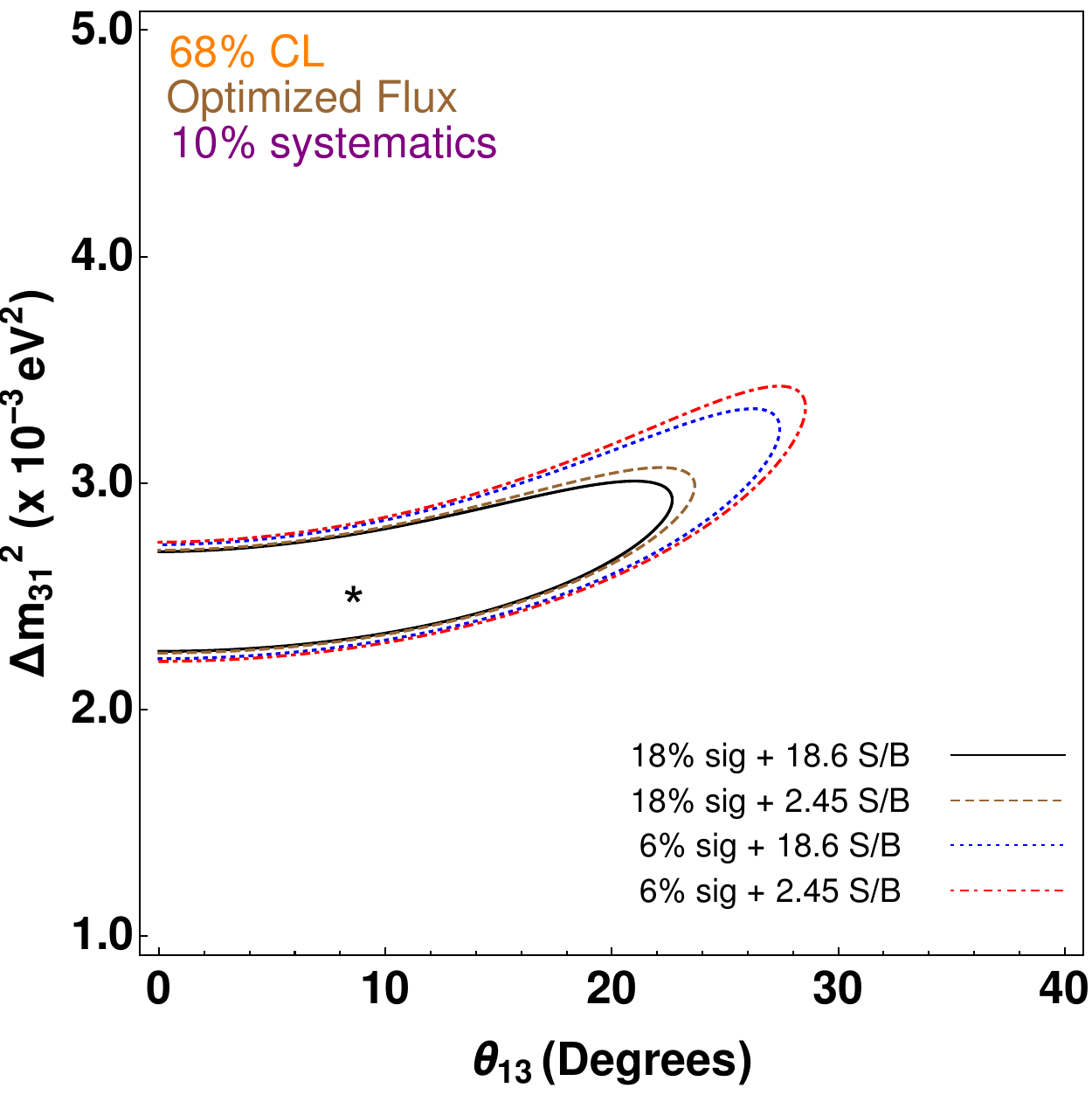} 
\includegraphics[height=7.5cm,width=7.5cm]{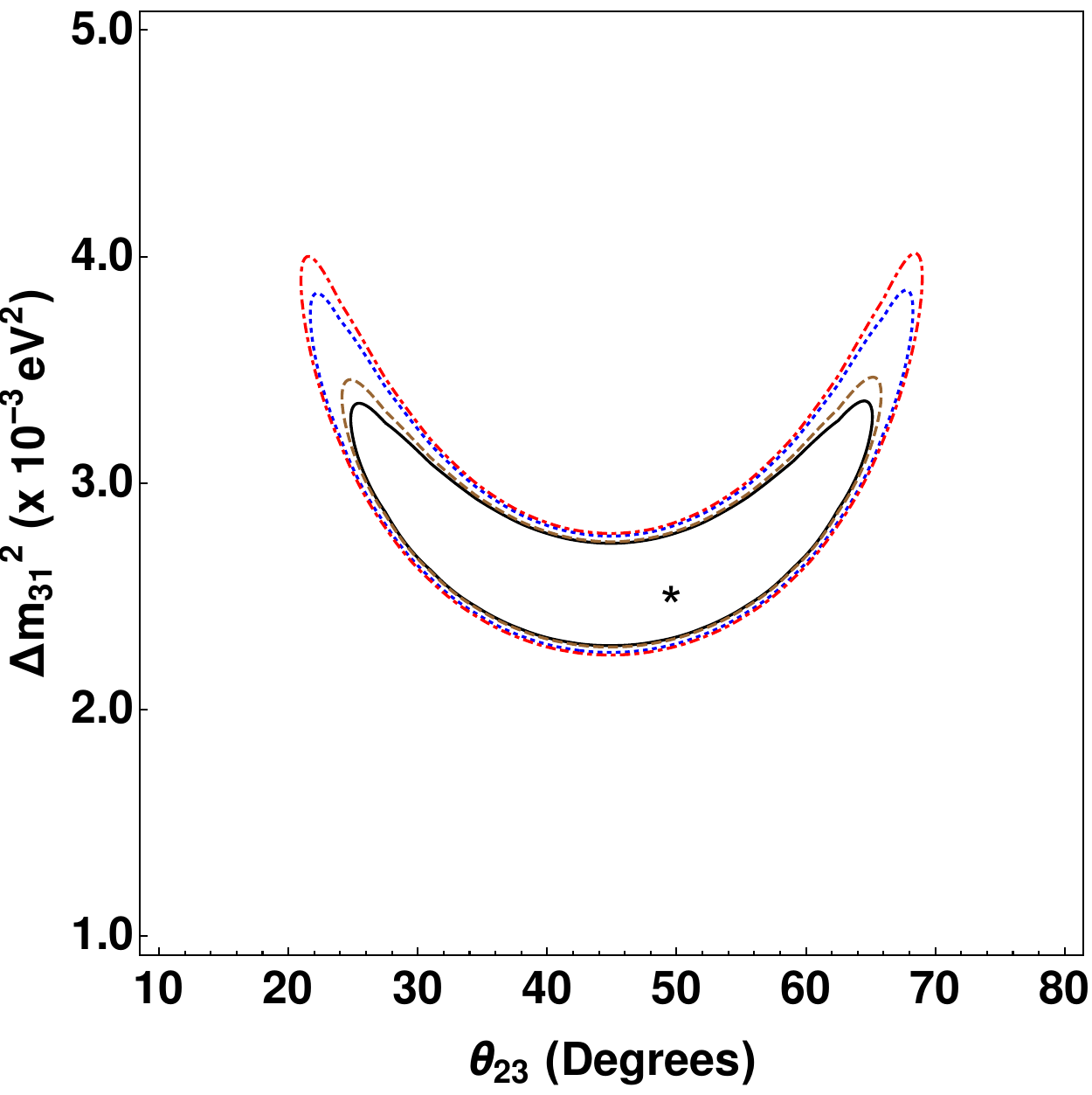} 
\includegraphics[height=7.5cm,width=7.5cm]{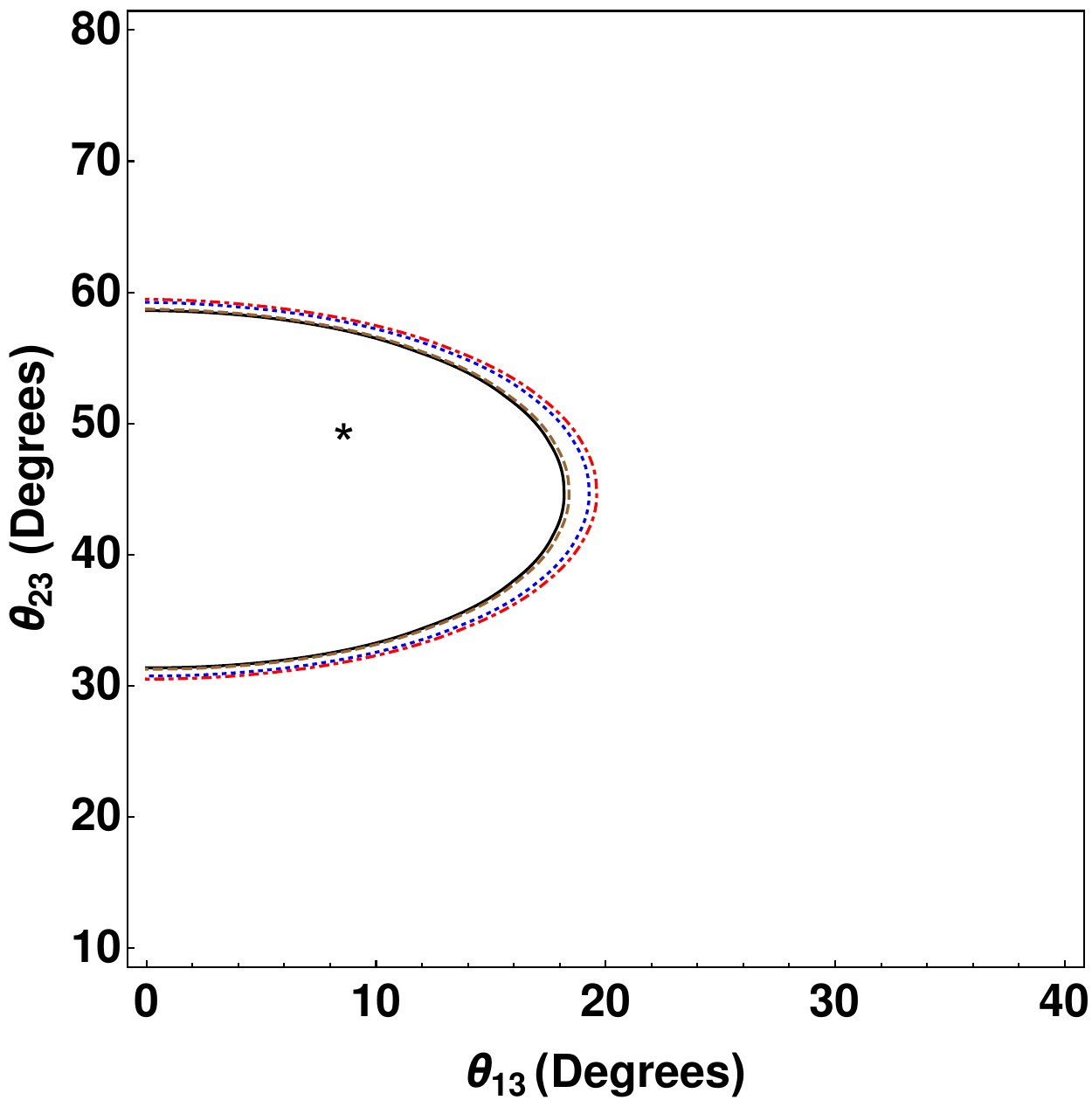} 
\includegraphics[height=7.5cm,width=7.5cm]{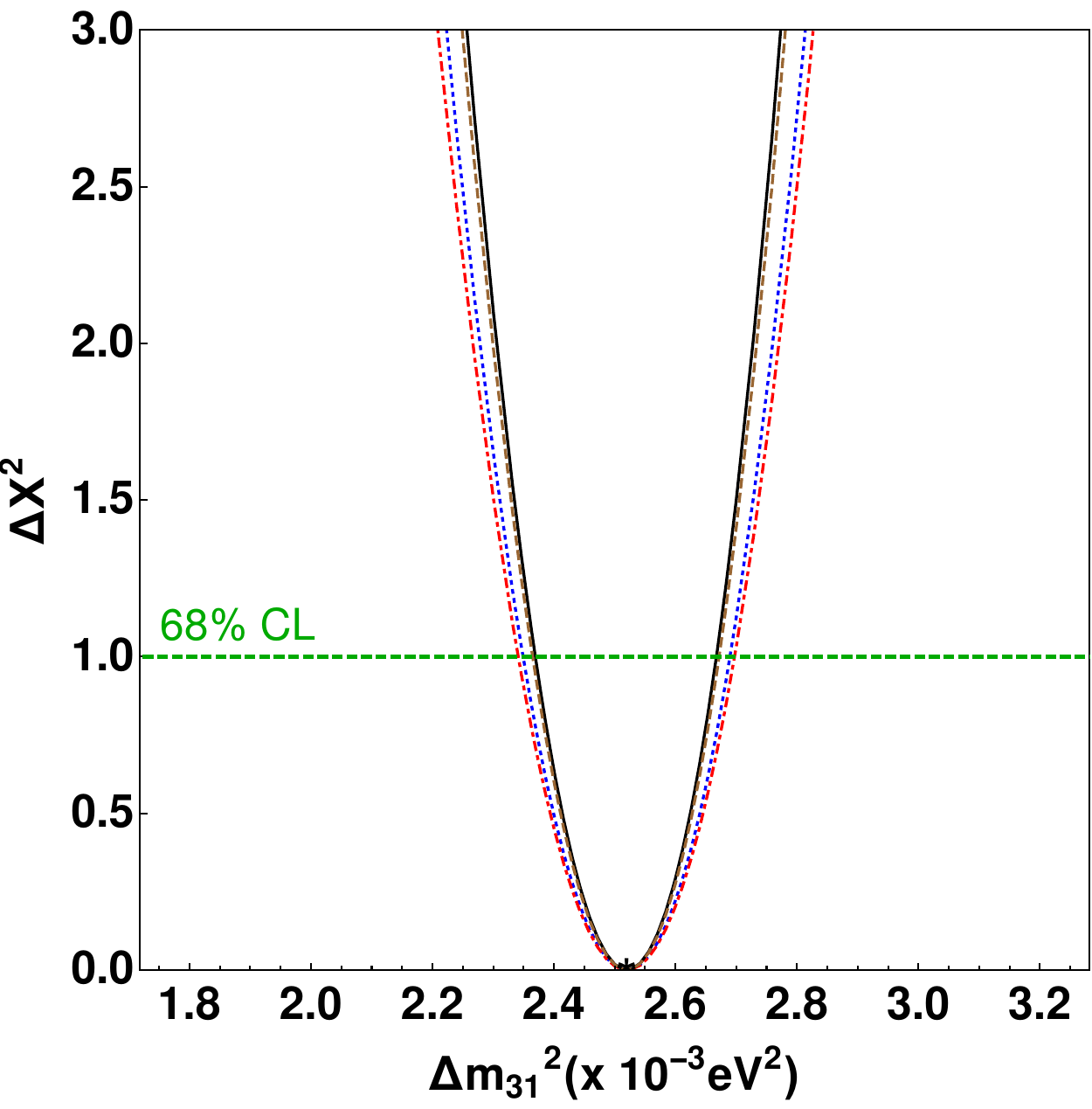}
\caption{\it Expected contours at 68\% CL in the oscillation parameter planes (as mentioned in each graph axis) for $S/B = 2.45$, with $\nu_{\tau}$ detection efficiency
of 6\% (Red, DotDashed) and 18\% (Brown, Dashed) and for $S/B = 18.6$ with $\nu_{\tau}$ detection efficiency of 6\% (Blue, Dotted) and 18\% (Black, Solid).
The tau optimized flux has been assumed in the simulation, using only the $\nu_{\tau}$ appearance channel in the Normal Hierarchy case with a  10\% signal uncertainty. 
Marginalization over all undisplayed parameters has been performed.
For the bottom right graph, $\Delta {\chi}^2 = \chi^2 -\chi_{min} ^2$ as a function of the true $\Delta m^2 _{31}$ is plotted.  
Stars represent the simulated true values.
} 
\label{sp_opt_10}
\end{center}
\end{figure}

From figure \ref{sp_opt_10} we see that an increase in the tau detection efficiency or even in the $S/B$ ratio impacts the parameters sensitivities less than in the case of the standard flux, the reason being that the flux is already optimized for tau as to produce a larger statistics in all cases.
Analizing the correlation plots, we observe that the smallest allowed ranges found in the case of the standard flux can be further reduced up to 10\% for $\theta_{23}$ and 15\% for $\theta_{13}$ if the optimized flux is considered. 
Notice also that with such a flux, DUNE can reach a relative uncertainty of 4.5\% in the measurement of the atmospheric mass difference if 10\% systematics on the signal is assumed.

\begin{figure}[tbp]
\begin{center}
\includegraphics[height=7.5cm,width=7.5cm]{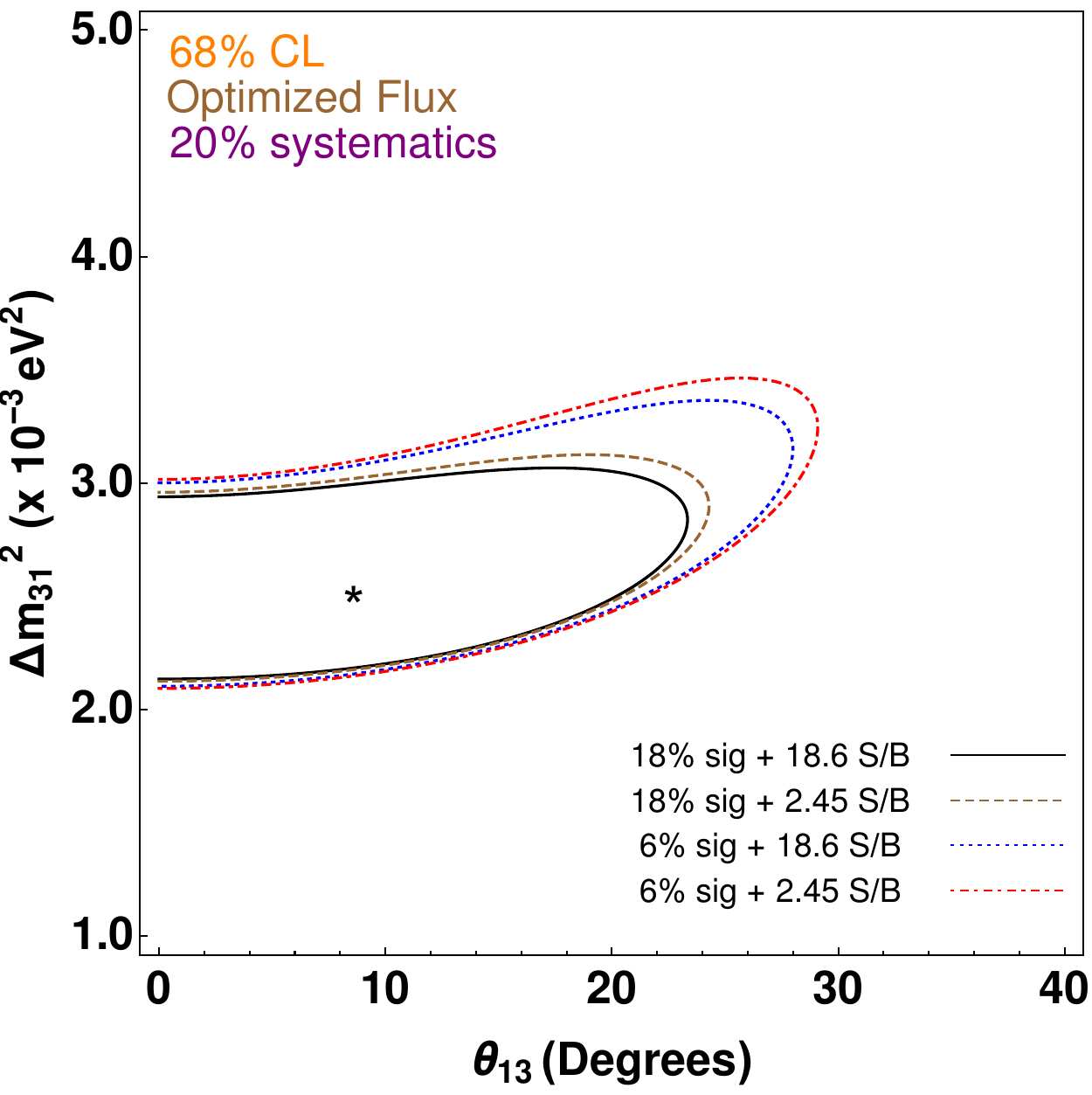} 
\includegraphics[height=7.5cm,width=7.5cm]{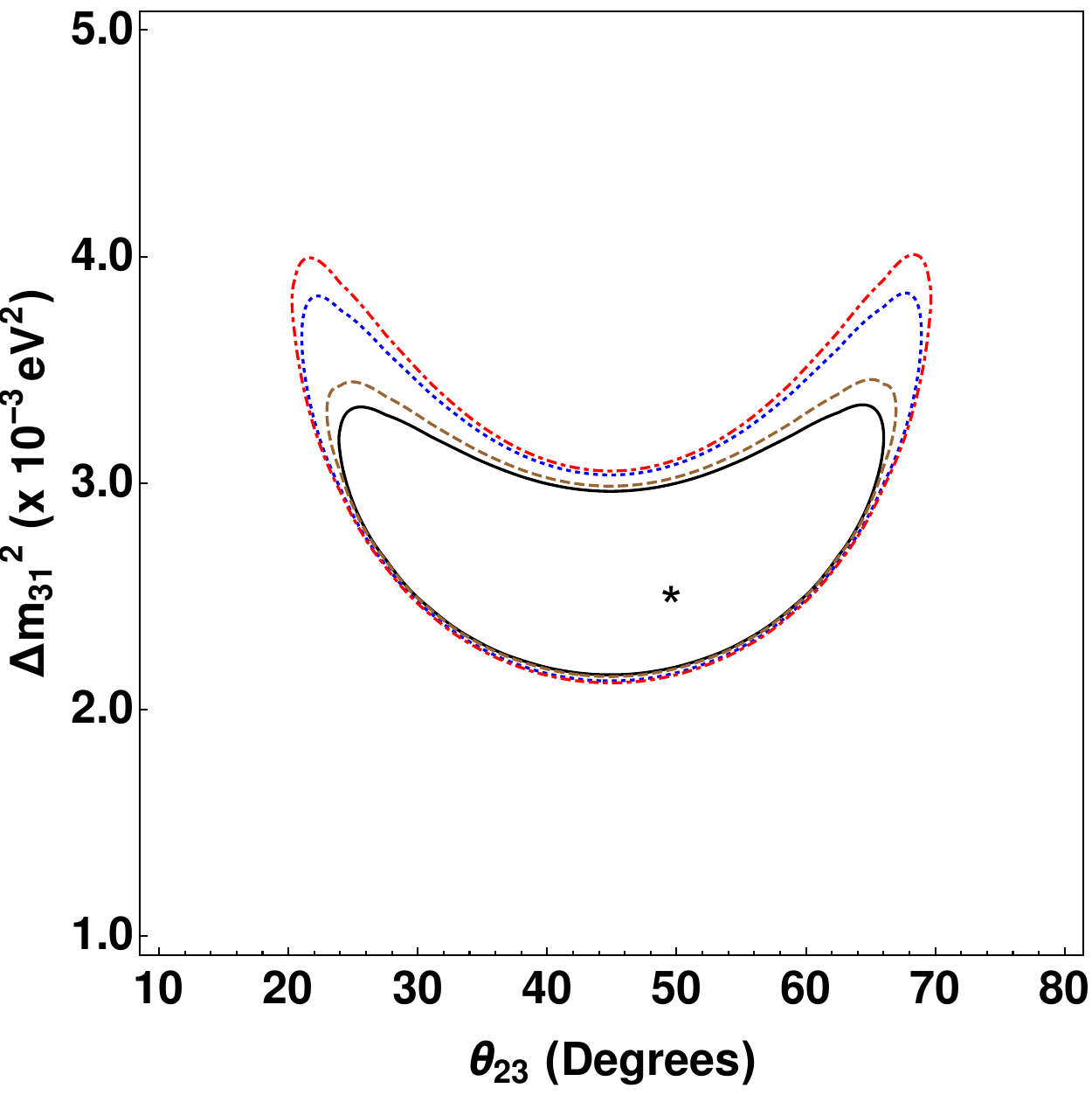} 
\includegraphics[height=7.5cm,width=7.5cm]{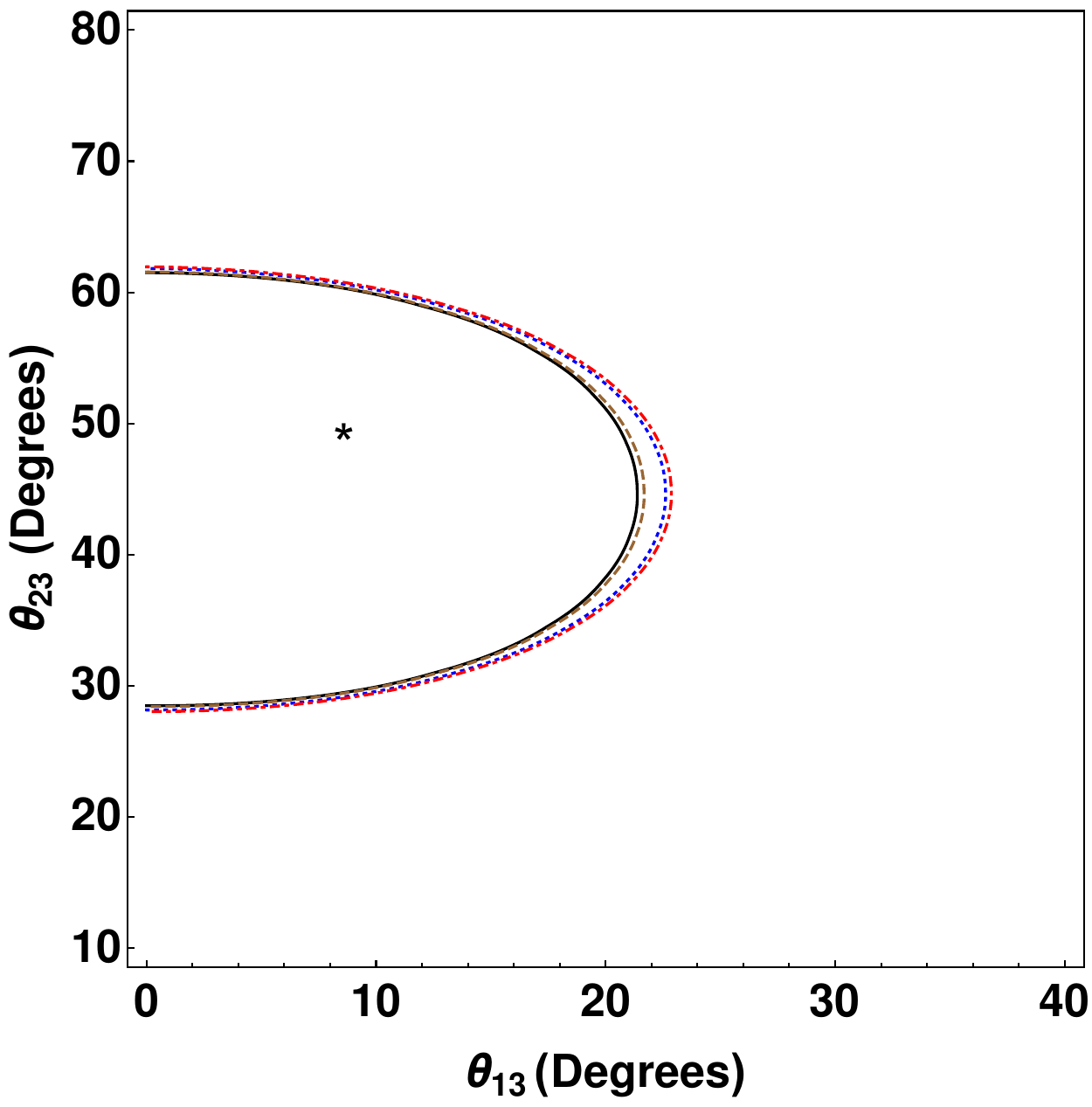} 
\includegraphics[height=7.5cm,width=7.5cm]{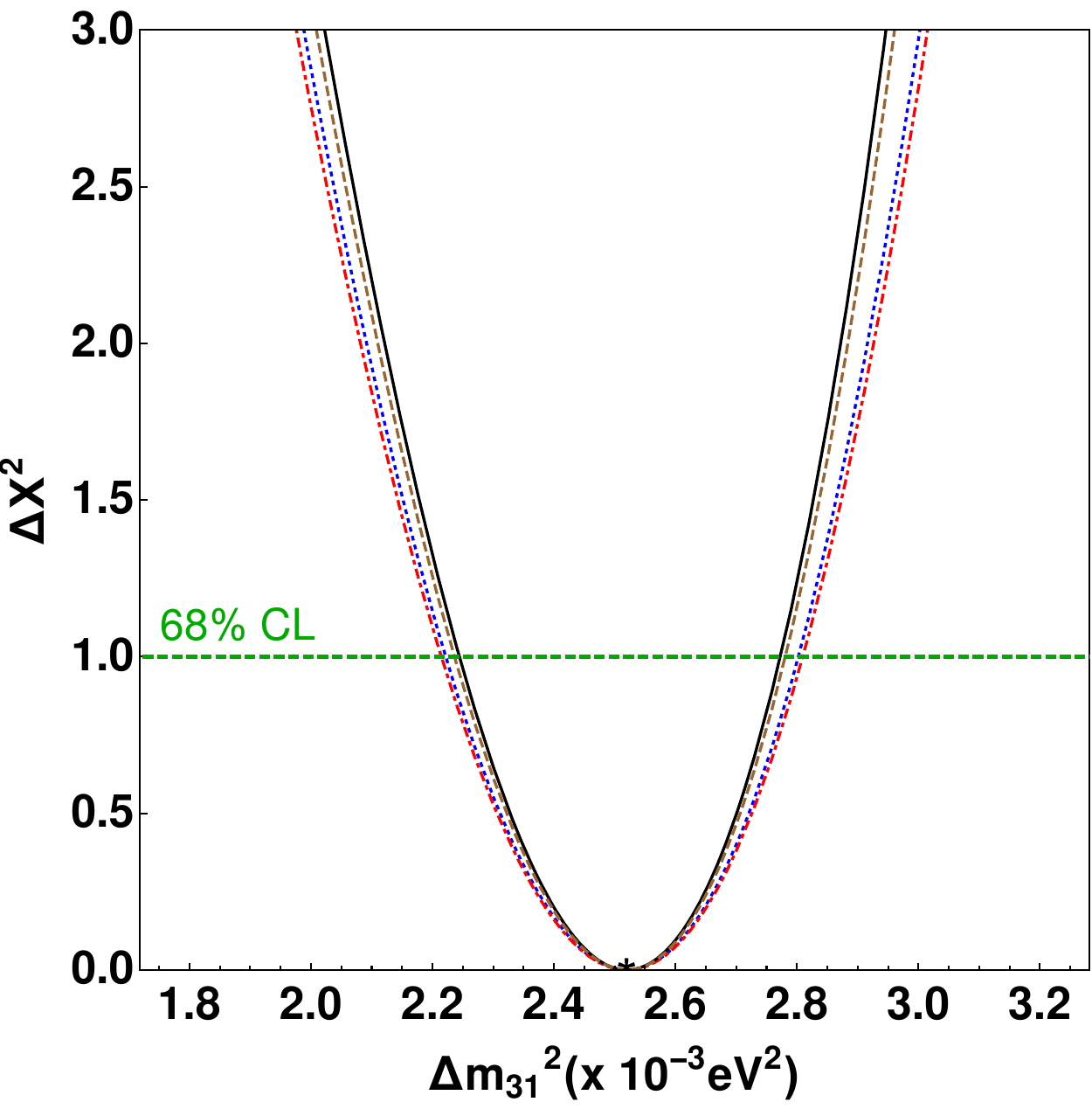} 
\caption{\it   Same as figure \ref{sp_opt_10} but with 20\% signal uncertainty.
} 
\label{sp_opt_20}
\end{center}
\end{figure}

Finally, in figure \ref{sp_opt_20} we present the results obtained for a 20\% signal uncertainty. As before, the improvement in the  sensitivity is smaller than in the case of the standard flux. The parameter which is affected the most by the systematics is the atmospheric mass splitting whose relative uncertainty is roughly 10\%.

\section{The case of Non-Standard Interaction}
\label{sect:nsi}
Several neutrino experiments have led to the opening for several new physics scenarios
among which NSIs are quite popular. 
If the SM is regarded as a low-energy effective theory of some higher theory in the UV, then BSM would enter as higher-dimensional
operators, suppressed by the new physics scale. 
%
In neutrino physics these are often written as four-fermion interactions, described by an effective four fermion Lagrangian~\cite{Davidson:2003ha}:
\begin{equation} 
-{\cal L}^{eff}_{\rm NSI} =
\varepsilon_{\alpha \beta}^{fP}{2\sqrt2 G_F} (\bar{\nu}_\alpha \gamma_\rho L \nu_\beta)
( \bar {f} \gamma^\rho P f ) \,,
\label{eq:efflag}
\end{equation}
where $G_F$ is the Fermi constant, $\varepsilon_{\alpha \beta}^{fP}$
is the parameter which describes the strength of the NSI, $f$ is a
first generation SM fermion ($e, u$ or $d$), $P$ denotes the chiral
projector $\{L,R=(1\pm\gamma^5)/2\}$, and $\alpha$ and $\beta$ denote
the neutrino flavors $e$, $\mu$ or $\tau$.


What is relevant for neutrino propagation in matter is the vector part $V$ of interaction 
 $\varepsilon_{\alpha\beta}^{f V} =\varepsilon_{\alpha\beta}^{f L}+\varepsilon_{\alpha\beta}^{f R}$ since the neutrino propagation in a medium is sensitive
to the combination $\varepsilon_{\alpha\beta} =\varepsilon_{\alpha\beta}^{e V}+ N_u/N_e \,\varepsilon_{\alpha\beta}^{u V}+N_d/N_e \,\varepsilon_{\alpha\beta}^{d V}$.
Following what is usually done in the literature, we discuss the bounds from DUNE  in terms of $\varepsilon_{\alpha\beta}$.
Overall, this Lagrangian describes
neutral current (NC) interactions; it basically modifies the matter Hamiltonian and consequently the 
transition probability of the neutrinos in matter. The strength of the new interaction is parameterized in terms of the complex 
$\varepsilon_{\alpha \beta}= |\epsilon_{\alpha\beta}| e^{i \phi_{\alpha\beta}}$
couplings. Thus the state evolution equations are given by: 
%
\begin{eqnarray} 
i \frac{d}{dt} \left( \begin{array}{c} 
                   \nu_e \\ \nu_\mu \\ \nu_\tau 
                   \end{array}  \right)
 = \left[ U_{PMNS} \left( \begin{array}{ccc}
                   0   & 0          & 0   \\
                   0   & \Delta_{21}  & 0  \\
                   0   & 0           &  \Delta_{31}  
                   \end{array} \right) U_{PMNS}^{\dagger} +  
                  A \left( \begin{array}{ccc}
            1 + \epsilon_{ee}     & \epsilon_{e\mu} & \epsilon_{e\tau} \\
            \epsilon_{e \mu }^*  & \epsilon_{\mu\mu}  & \epsilon_{\mu\tau} \\
            \epsilon_{e \tau}^* & \epsilon_{\mu \tau }^* & \epsilon_{\tau\tau} 
                   \end{array} 
                   \right) \right] ~
\left( \begin{array}{c} 
                   \nu_e \\ \nu_\mu \\ \nu_\tau 
                   \end{array}  \right)\, ,
\label{eq:matter}
\end{eqnarray}
where  $\Delta_{ij}=\Delta m^2_{ij}/2E$, $U_{PMNS}$ is the neutrino mixing matrix,
$A\equiv 2 \sqrt 2 G_F n_e$ with $n_e$ being the electron density in the Earth crust. All in all, 
beside the standard oscillation parameters, the parameter space is enriched by six more moduli $|\epsilon_{\alpha\beta}|$ 
and three more phases $\phi_{\alpha\beta}$. 

Direct constraints on NSI can be derived from scattering processes \cite{Biggio:2009nt} and from neutrino oscillation data \cite{Gonzalez-Garcia:2013usa}.
Latest constrains on NSI parameters from global fits were quoted in \cite{Esteban:2019lfo}, from which 
we extracted the limits reported in table \ref{tab:limits}, used in our numerical analysis.  
\begin{table}[h]
\begin{center}
\begin{tabular}{ |c|c| }
\hline
NSI parameters &   Limits \\ \hline
$\epsilon_{e e} - \epsilon_{\mu \mu}$ & (-0.2, 0.45) \\
$|\epsilon_{e \mu}|$ & $<$  0.1 \\
$|\epsilon_{e \tau}|$ & $<$ 0.3 \\
$\epsilon_{\tau\tau} - \epsilon_{\mu\mu}$ & (-0.02, 0.175) \\
$|\epsilon_{\mu\tau}|$ &  $<$ 0.03\\ \hline
\end{tabular}
\caption{\it Current constraints on the NSI parameters at 90\% CL obtained from a global fit to
neutrino oscillation data \cite{Esteban:2019lfo}. No bounds on the phases $\phi_{\alpha\beta}$ are available so far.}
\label{tab:limits}
\end{center}
\end{table}
We want to outline that, in order to compute the oscillation probabilities in presence of NSI, we may subtract from the diagonal entries any one of the diagonal elements $\epsilon_{\alpha\alpha}$
as the oscillation phenomenon is insensitive to overall factors.
Therefore one may consider, as done in table \ref{tab:limits}, the {\it shifted} parameters 
$\epsilon_{ee} - \epsilon_{\mu\mu}$ and $\epsilon_{\tau\tau} - \epsilon_{\mu\mu}$ instead of $\epsilon_{ee}$ and $\epsilon_{\tau\tau}$ respectively and set $\epsilon_{\mu\mu}$ to 0,
which is also motivated by the strong external bounds  as well \cite{Ohlsson:2012kf}.

\subsection{The importance of \texorpdfstring{$\nu_\mu \to \nu_\tau$}{numu to nutau} channel}

General NSI considerations were already investigated in Refs. \cite{deGouvea:2015ndi,Coloma:2015kiu} and the effect of systematics were studied in \cite{Meloni:2018xnk} 
in the standard DUNE scenario consisting of the $\nu_e$ appearance and $\nu_{\mu}$ disappearance channels.  
  The $\nu_\mu \rightarrow \nu_e$  oscillation probability is affected by 
the $\epsilon_{e\mu}$ and $\epsilon_{e\tau}$ parameters, as well as by $\epsilon_{ee}$. However, statistics in DUNE is dominated by the disappearance channel $\nu_\mu \rightarrow \nu_\mu$  which is mainly affected by the presence of 
  $\epsilon_{\tau\tau}$ and $\epsilon_{\mu\tau}$. The dependence of probabilities on these parameters have been studied, among others, in
  \cite{Kopp:2007ne,Kikuchi:2008vq,Blennow:2008ym}.

For the appearance probability $\nu_\mu \rightarrow \nu_{\tau}$, the leading analytic behavior in terms of the small $\epsilon$ parameters is given by \cite{Kikuchi:2008vq}:
\begin{equation}
\label{eq:pmutau}
P_{\mu\tau}  = P_{\mu\tau}^{SM} + \left(\frac{1}{2} \epsilon_{\tau\tau} \cos ^2 (2\theta_{23}) + 2\cos (2\theta_{23}) \text{Re}\{\epsilon_{\mu\tau} \}\right) 
\left( AL\right)\sin\left(\frac{\Delta m ^{2} _{31}L}{2E}\right)
+ {\cal O}(\epsilon ^2),
\end{equation}
where we neglected the small solar mass squared difference $\Delta m_{21}^2$; $P_{\mu\tau}^{SM}$ is the oscillation 
probability in absence of NSI already discussed in section \ref{stphys}. Additional terms, which depend on both the real and imaginary parts of $\epsilon_{\mu\tau}$, are at second order in the 
perturbative expansion, and are expected to provide only small sensitivity in the regions with $\phi_{\mu\tau} \sim \pm \pi/2$. Since the term containing $\epsilon_{\tau\tau}$ is  very small for an almost maximal atmospheric angle, the probability $P_{\mu \tau}$ is  sensitive to large values of $\epsilon_{\tau\tau}$ only, unlike for $\epsilon_{\mu\tau}$ to which we expect a maximal sensitivity.

\subsection{Effect of Systematics, Experimental Reach and \texorpdfstring{$\nu_{\tau}$}{nutau} Detection Efficiency}

\begin{figure}[b]
\begin{center}
\includegraphics[height=7.5cm,width=7.5cm]{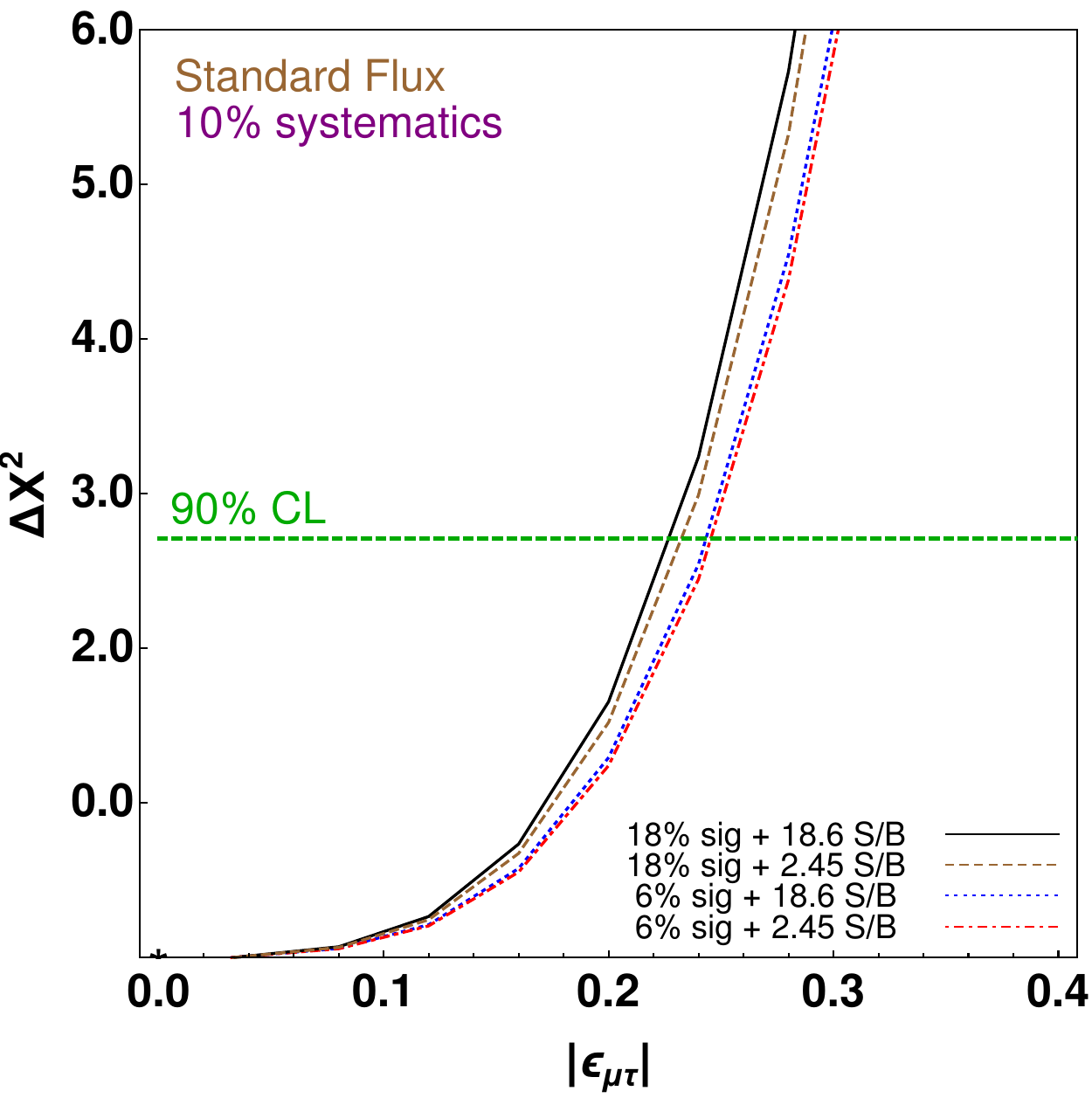} 
\includegraphics[height=7.5cm,width=7.5cm]{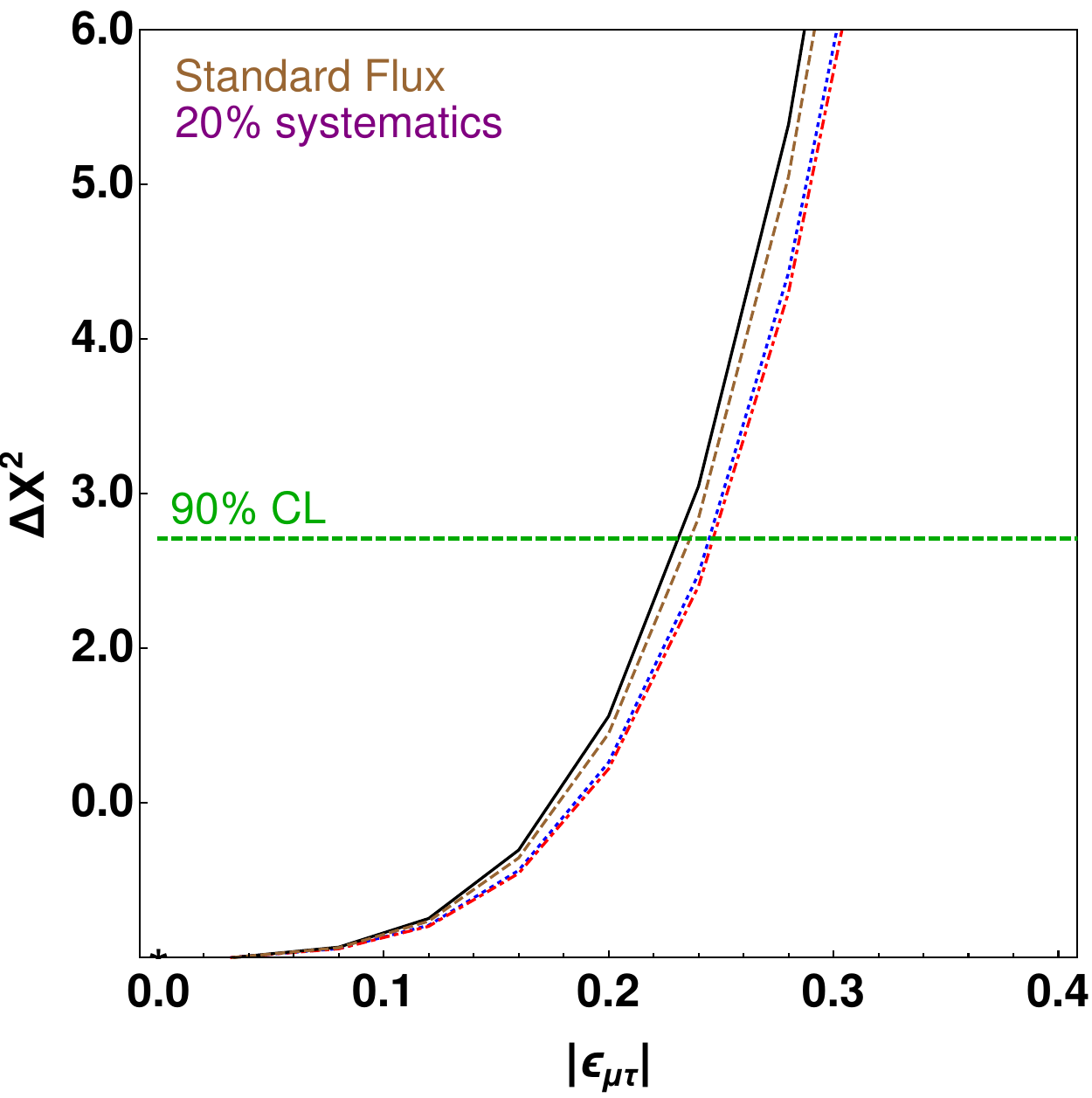} 
\caption{\it $\Delta \chi^2$ vs $|\epsilon_{\mu\tau}|$ for 10\% (left panel) and  20\% (right panel) $\nu_\tau$ signal systematic uncertainty. The standard neutrino flux has been assumed in the simulations. Horizontal dashed line indicates the 90 \% CL limit (1 degree of freedom). The meaning of the curves is the same as the previous plots.
}
\label{fig:chi_mutau_std_10sys}
\end{center}
\end{figure}

\begin{figure}[h]
\begin{center}
\includegraphics[height=7.5cm,width=7.5cm]{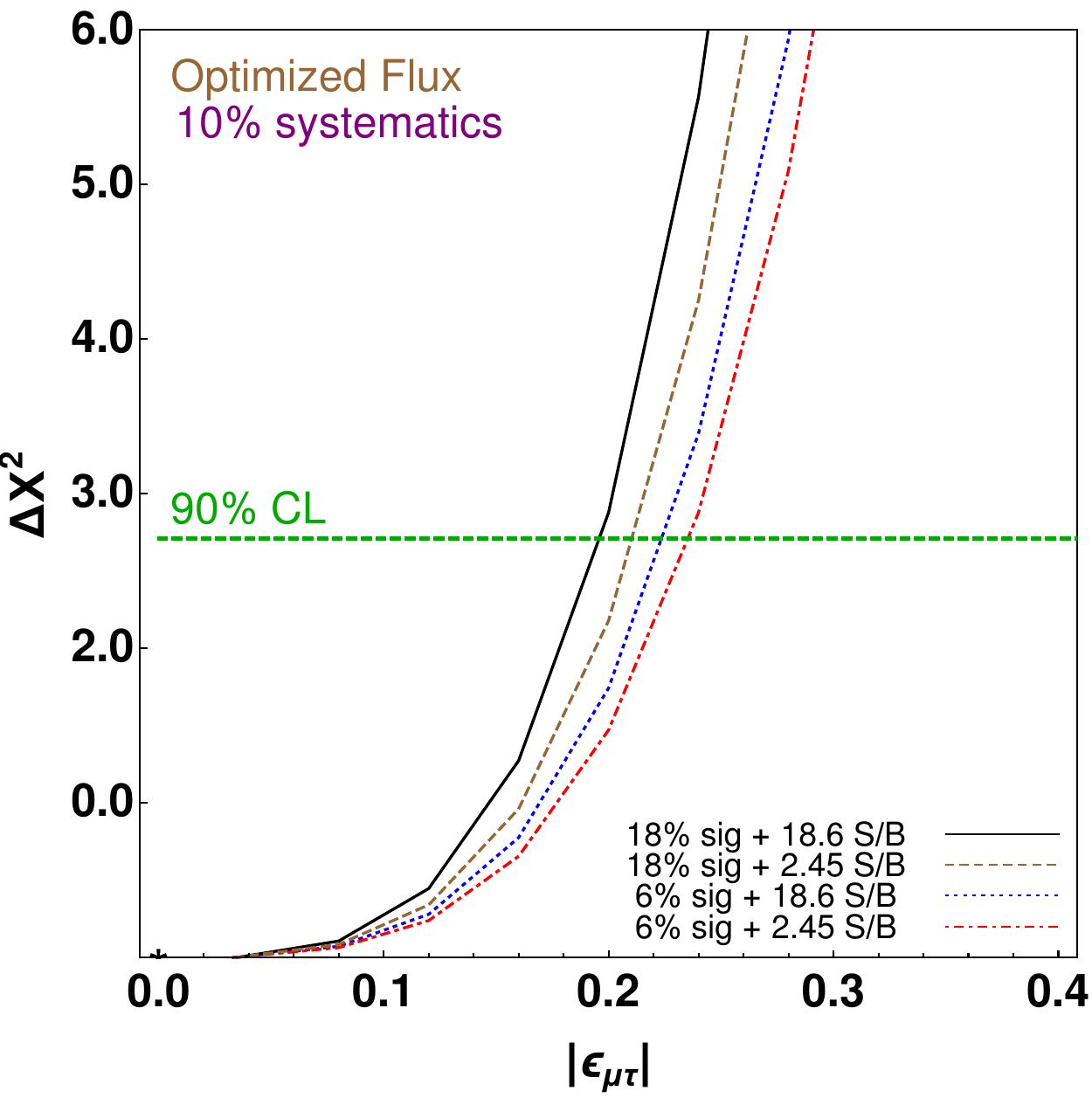}  
\includegraphics[height=7.5cm,width=7.5cm]{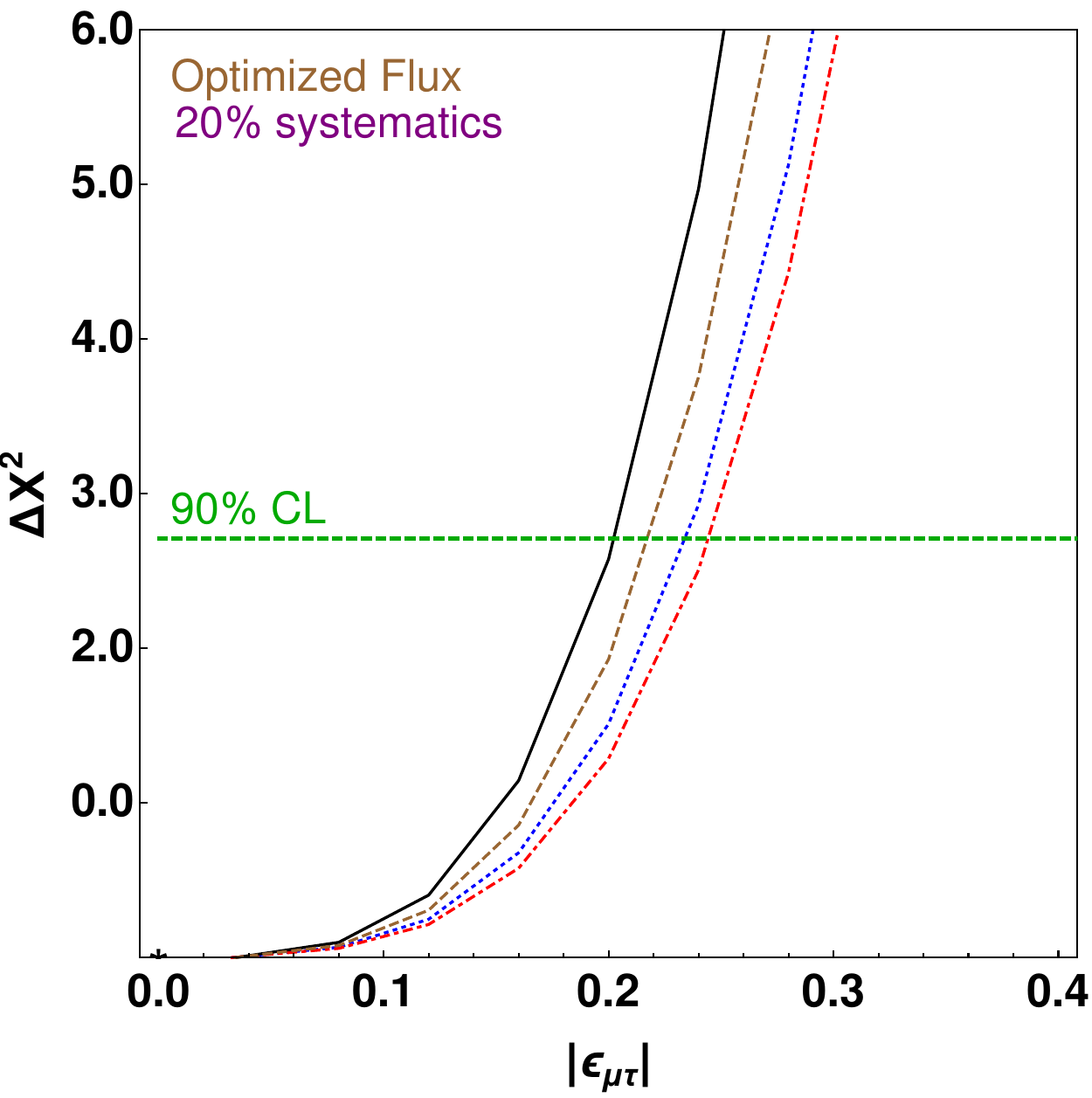} 
\caption{\it Same as figure \ref{fig:chi_mutau_std_10sys} but for the optimized flux. 
}
\label{fig:chi_mutau_opt_10sys}
\end{center}
\end{figure}

In figures \ref{fig:chi_mutau_std_10sys} and \ref{fig:chi_mutau_opt_10sys} we report 
the $\Delta \chi^2$ as a function of  $|\epsilon_{\mu\tau}|$ for 10\% (left panel) and  20\% (right panel) systematic uncertainties  for the $\nu_\tau$ signal, for the standard and optimized fluxes, respectively. 
Plots have been obtained marginalizing over all standard and
non standard parameters according to the constraints reported in tables 2 and
5, except for the solar parameters $\theta_{12}$ and $\Delta m^2_{21}$ and the NSI parameter $\varepsilon_{\mu\tau}$, which
have been fixed to their best fit values.
All the NSI phases have been left free with no bounds.

The common feature of the above figures is that there is a significant improvement in the bound for $|\epsilon_{\mu\tau}|$ by increasing the efficiency from 6\% upto 18\%, and the S/B from 2.45 upto 18.6;   one can envisage an overall improvement at 90\% CL of approximately 10\% in the parameter relative uncertainty in the case of the  standard flux and 18\% using the
optimized flux.

Sensitivity limits at $90\%$ CL reached with all the three appearance and disappearance channels in DUNE using the two fluxes are presented in the Tables \ref{fig:Range_constrained_10sys_stdflux} and  \ref{fig:Range_constrained_20sys_stdflux}.
The worst case scenario using the standard flux gives us a limit on the $|\epsilon_{\mu\tau}| \lesssim 0.25$ while the most stringent limit can be set using the optimized flux with the best efficiency ($18\%$), the best 
\begin{table}[h]
\begin{center}
\begin{tabular}{c|c|c|c|c|c|c|c|c|}
\cline{2-5}
                                                           & \multicolumn{4}{c|}{\textbf{Standard Flux ($10\%$ sys)}}                                                                                                                                                  \\ \cline{2-5} 
                                                           & \multicolumn{2}{c|}{S/B = 2.45}                    & \multicolumn{2}{c|}{S/B = 18.6}                                        \\ \hline
\multicolumn{1}{|c|}{Efficiency of $\nu_{\tau}$ detection} & $6 \%$                 & $18 \%$                             & $6 \%$                  & $18 \%$                                                       \\ \hline                                              
\multicolumn{1}{|c|}{$|\epsilon_{\mu\tau}|$}                 & [0,0.2452]             &  [0,0.2320]                         & [0,0.2431]              & [0,0.2264]                                                       \\ \hline
& \multicolumn{4}{c|}{\textbf{Optimized Flux ($10\%$ sys)}}                                                                     \\ \cline{2-5} 
                                                             & \multicolumn{2}{c|}{S/B = 2.45}       & \multicolumn{2}{c|}{S/B = 18.6}                        \\ \hline
\multicolumn{1}{|c|}{Efficiency of $\nu_{\tau}$ detection}   & $6 \%$                  & $18 \%$               & $6 \%$                  & $18 \%$                                \\ \hline                                              
\multicolumn{1}{|c|}{$|\epsilon_{\mu\tau}|$}                   & [0,0.2349]              & [0,0.2101]             & [0,0.2232]               & [0,0.1955]                                \\ \hline

\end{tabular}
\caption{\it  Summary of 90\% CL bounds on NSI parameter $|\epsilon_{\mu\tau}|$ that DUNE may set for 10\% systematic uncertainty for the $\nu_{\tau}$ appearance channel.} 
\label{fig:Range_constrained_10sys_stdflux}
\end{center}
\end{table}

\begin{table}[h]
\begin{center}
\begin{tabular}{c|c|c|c|c|c|c|c|c|}
\cline{2-5}
                                                           & \multicolumn{4}{c|}{\textbf{Standard Flux ($20\%$ sys)}}                                                                                                                                                   \\ \cline{2-5} 
                                                           & \multicolumn{2}{c|}{S/B = 2.45}                    & \multicolumn{2}{c|}{S/B = 18.6}                                                 \\ \hline
\multicolumn{1}{|c|}{Efficiency of $\nu_{\tau}$ detection} & $6 \%$                 & $18 \%$                             & $6 \%$                  & $18 \%$                                                          \\ \hline                                              
\multicolumn{1}{|c|}{$|\epsilon_{\mu\tau}|$}                 & [0,0.2463]             &  [0,0.2359]                         & [0,0.2445]               & [0,0.2306]                                                        \\ \hline
& \multicolumn{4}{c|}{\textbf{Optimized Flux ($20\%$ sys)}}                                                                     \\ \cline{2-5} 
                                                                  & \multicolumn{2}{c|}{S/B = 2.45}       & \multicolumn{2}{c|}{S/B = 18.6}                        \\ \hline
\multicolumn{1}{|c|}{Efficiency of $\nu_{\tau}$ detection}        & $6 \%$                  & $18 \%$               & $6 \%$                  & $18 \%$                                \\ \hline                            

\multicolumn{1}{|c|}{$|\epsilon_{\mu\tau}|$}                        & [0,0.2440]              & [0,0.2169]             & [0,0.2335]               & [0,0.2021]                               \\ \hline

\end{tabular}
\caption{\it  Same as table \ref{fig:Range_constrained_10sys_stdflux} but for a 20\% systematic uncertainty for the $\nu_{\tau}$ appearance channel.} 
\label{fig:Range_constrained_20sys_stdflux}
\end{center}
\end{table}
S/B (18.6) and $10\%$ systematic uncertainty, $|\epsilon_{\mu\tau}| \lesssim 0.20$. 
This limit is approximately $35\%$ smaller than 
the one that can be set by DUNE using only $\nu_{e}$ appearance and $\nu_{\mu}$ disappearance channels with standard flux, 
$|\epsilon_{\mu\tau}| < 0.32$, as estimated in Ref. \cite{Meloni:2018xnk}.
We observe that a fit to the OPERA $\nu_{\tau}$ events \cite{Agafonova:2018auq} did in Ref. \cite{Meloni:2019pse}  predicted the bound
$|\epsilon_{\mu\tau}| \lesssim 0.41$ (marginalising over all parameters including NSI parameters) which is almost a factor of two larger than the worst limit DUNE can set.


We checked that the other NSI parameters do not benefit so much from the $\nu_\tau$ appearance channel and the sensitivity reach remains roughly
the same as in the standard DUNE scenario with $\nu_e$ appearance and $\nu_\mu$ disappearance only.

\section{The case of 3+1}
\label{sec:sterile}

The 3+1 sterile neutrino scenario is a possible solution to the short-baseline anomalies, which are about an excess of oscillations in short baseline experiments. For this reason, this new physics model is one of the most studied so far, and it is expected to be tested more in future experiments.
The 3+1 sterile neutrino model is based on the addition of a fourth mass-eigenstate, $m_4$, to the other standard three. Furthermore, the new interaction eigenstate is assumed to be sterile, that is not involved in electro-weak interactions with matter.

Adding a new neutrino to the standard model introduces  new parameters (three more mixing angles, $\theta_{14}$, $\theta_{24}$ and $\theta_{34}$, two more phases $\delta_1$ and $\delta_3$ and a new independent mass-squared splitting $\Delta m_{41}^2$) whose sensitivity will be studied in the following, with particular emphasis to the constraints attainable from the $\nu_{\tau}$ appearance channel.
 
The $U_{PMNS}$ matrix in this case is a unitary $4\times 4$ matrix. The parameterization used here  is as follows \cite{Maltoni:2007zf,Donini:2007yf,Meloni:2010zr}:
\begin{equation}
    U_{PMNS}=R(\theta_{34}) \, R(\theta_{24}) \, R(\theta_{23}, \delta_2) \, R(\theta_{14}) \, R(\theta_{13}, \delta_3) \, R(\theta_{12}, \delta_1)\,,
\end{equation}
where the phase $\delta_3$ reduces to the standard $\delta_{CP}$ when the new mixing angles are set to zero.
In our numerical simulations no external priors on the $\theta_{i4}$'s have been considered.

\subsection{The importance of \texorpdfstring{$\nu_\mu \to \nu_\tau$}{numu to nutau} channel}

The oscillation probabilities $\nu_{\mu} \rightarrow \nu_{\tau} $ in the case of the sterile neutrino model are more complicated than the standard ones. However in the vacuum it is possible to make some useful approximations (matter effects have been taken into account in the numerical simulations).  

If  the new mass-squared splitting satisfies $|\Delta m_{41}^2| \gg|\Delta m_{32}^2|$, it is possible to average all the trigonometric functions which include $\Delta m_{41}^2$.  Using the additional approximation of vanishing $\Delta m_{21}^2$, the following $\nu_{\mu} \rightarrow \nu_{\tau} $ oscillation formula can be obtained:

\begin{equation}
\begin{split}
 P_{\mu\tau}=& 2|U_{\tau 4}|^2|U_{\mu 4}|^2+4 \Re[  U_{\mu 3}^*U_{\tau 3}(U_{\mu 3}U_{\tau 3}^*+U_{\mu 4}U_{\tau 4}^*)]     \sin^2 \left (  \frac{  \Delta m_{32}^2 L}{4E} \right )+  \\ 
 &  -2\Im ( U_{\mu 3}^*U_{\tau 3}U_{\mu 4}U_{\tau 4}^*)\sin \left (  \frac{ \Delta m_{32}^2 L}{2E} \right ) \,.
\end{split}
\label{mu-tau}
\end{equation}

On the other hand, in the regime where $|\Delta m_{41}^2| \sim |\Delta m_{21}^2|$, an expansion to the first order in the small parameters $\alpha = \frac{ \Delta m_{21}^2 L}{4E}$ and $\beta =  \frac{\Delta m_{41}^2 L}{4E}$ produces the following result:

\begin{equation}
\begin{split}
P_{\mu\tau} = & 4 \{ |U_{\mu 3}|^2|U_{\tau 3}|^2+\\ & \Im[\alpha U_{\mu 3}^*U_{\tau 3}U_{\mu 1}U_{\tau 1}^*+(\beta-\alpha)U_{\mu 4}^*U_{\tau 4}U_{\mu 3}U_{\tau 3}^*] \} sin^2 \left ( \frac{ \Delta m_{32}^{2} L}{4E} \right)+\\ 
 & -2\Re[\alpha U_{\mu 3}^*U_{\tau  3}U_{\mu 1}U_{\tau 1}^*-(\beta-\alpha)U_{\mu 4}^*U_{\tau 4}U_{\mu 3}U_{\tau 3}^*]sin\left (  \frac{\Delta m_{32}^{2} L}{2E} \right)\,.
\end{split}
\label{mu-taulight}
\end{equation}
Both eqns.(\ref{mu-tau}) and (\ref{mu-taulight}) show that the $\nu_{\mu} \rightarrow \nu_{\tau} $  probability is driven by the combination $U_{\mu 4}^*U_{\tau 4}$ of new mixing angles that, as we have verified, does not appear in any of the other oscillation probabilities at the leading order. In the parameterization used in this paper, such a  combination can be written in terms of mixing angles as:
\begin{equation}
    U_{\mu 4}^*U_{\tau 4} = \frac{1}{2}\cos^2\theta_{14}\sin\theta_{34}\sin2\theta_{24}.
\end{equation}
This relation shows that the relevant changes in $P_{\mu\tau}$ with respect to the three neutrino scenario are mainly due to $\theta_{34}$ and $\theta_{24}$.
Since the combinations $U_{\mu 4}^*U_{e 4}$ and $|U_{\mu 4}|^2$ that appear in the other oscillation probabilities contain $\theta_{24}$ as well \cite{Donini:2008wz,Klop:2014ima,Berryman:2015nua,Gandhi:2015xza}, we expect $\theta_{34}$ to be the parameter whose sensitivity will be mostly affected by the addition of the $\nu_{\tau}$ appearance channel.

\subsection{Constraints on Sterile Neutrino Parameters}

\begin{figure}[h]
\begin{center}
\includegraphics[height=7.5cm,width=7.5cm]{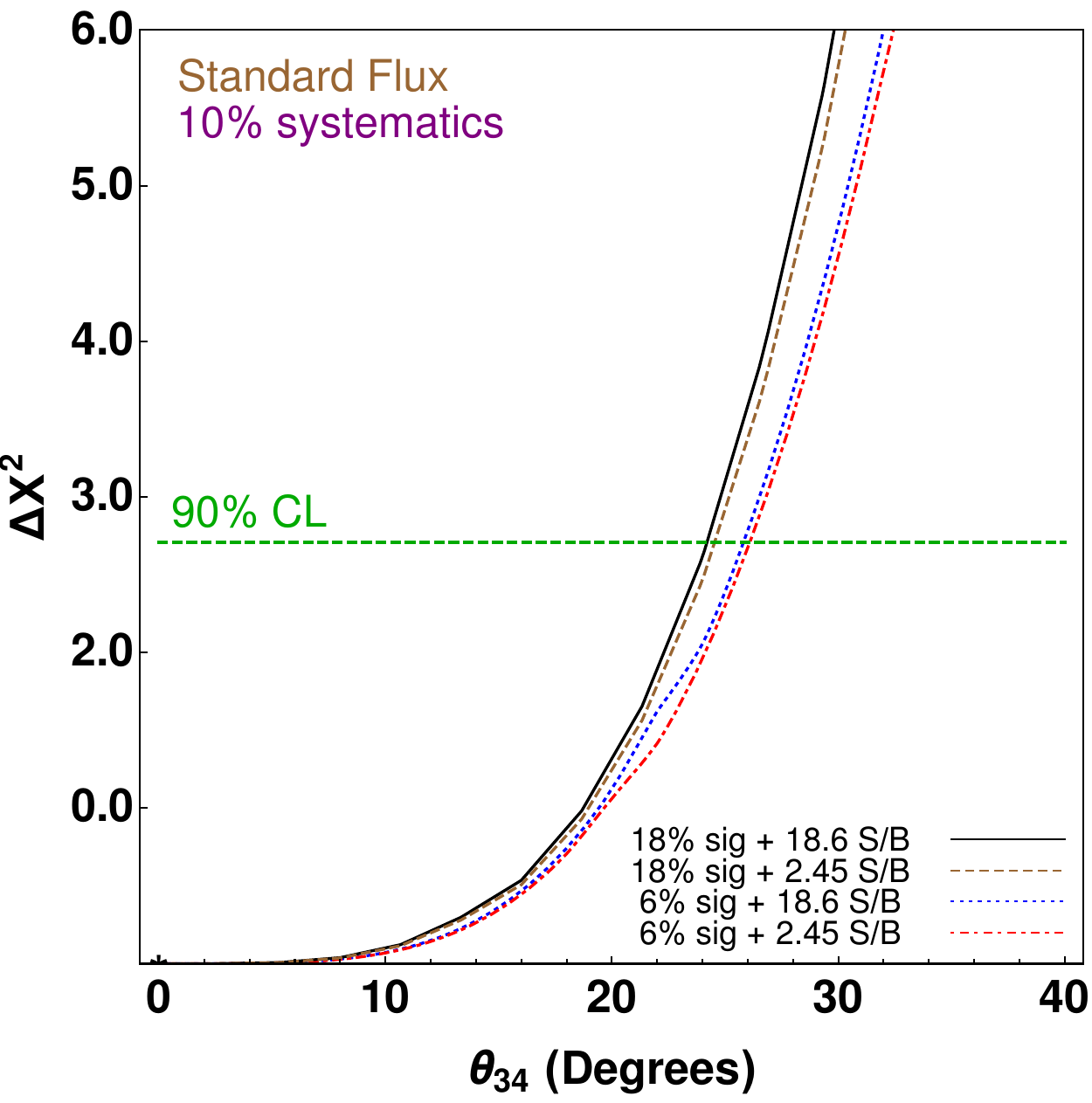} 
\includegraphics[height=7.5cm,width=7.5cm]{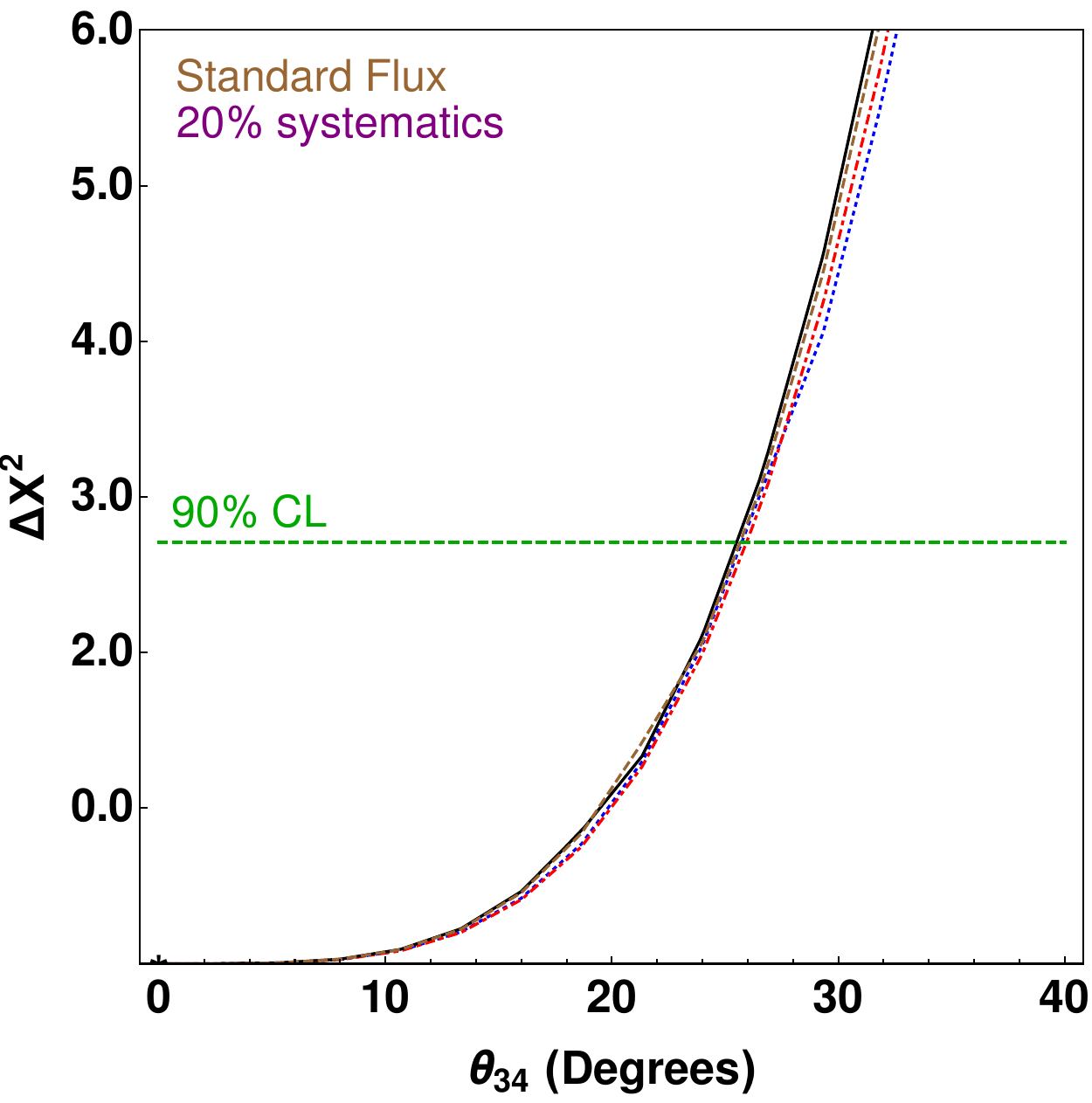} 
\caption{\it  Expected sensitivity to $\theta_{34}$ obtained assuming  the  standard flux and $\Delta m_{41}^2=1 \, eV^2$. See text for more details.
} 
\label{fig:sens34_std}
\end{center}
\end{figure}

\begin{figure}[h]
\begin{center}
\includegraphics[height=7.5cm,width=7.5cm]{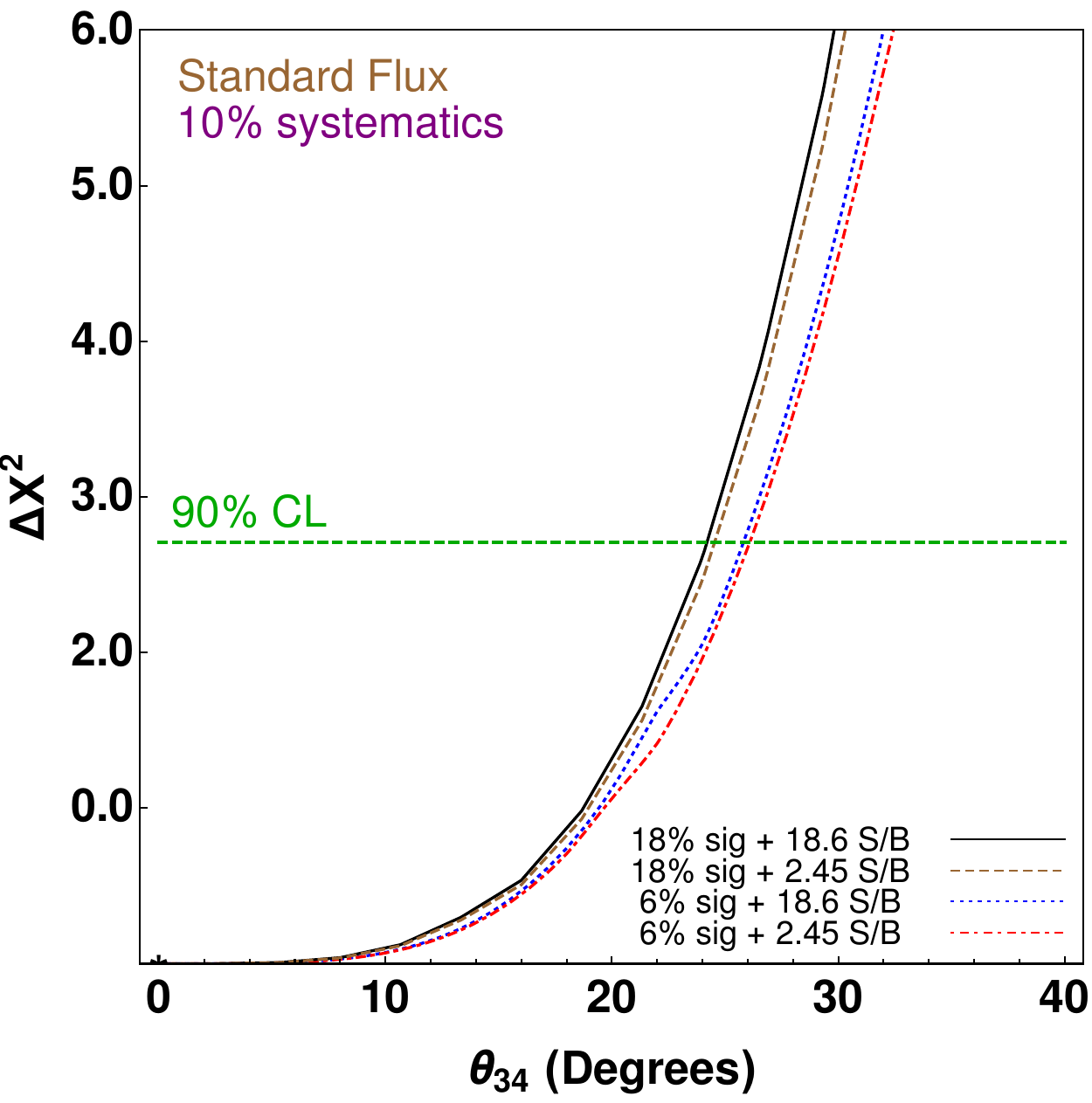} 
\includegraphics[height=7.5cm,width=7.5cm]{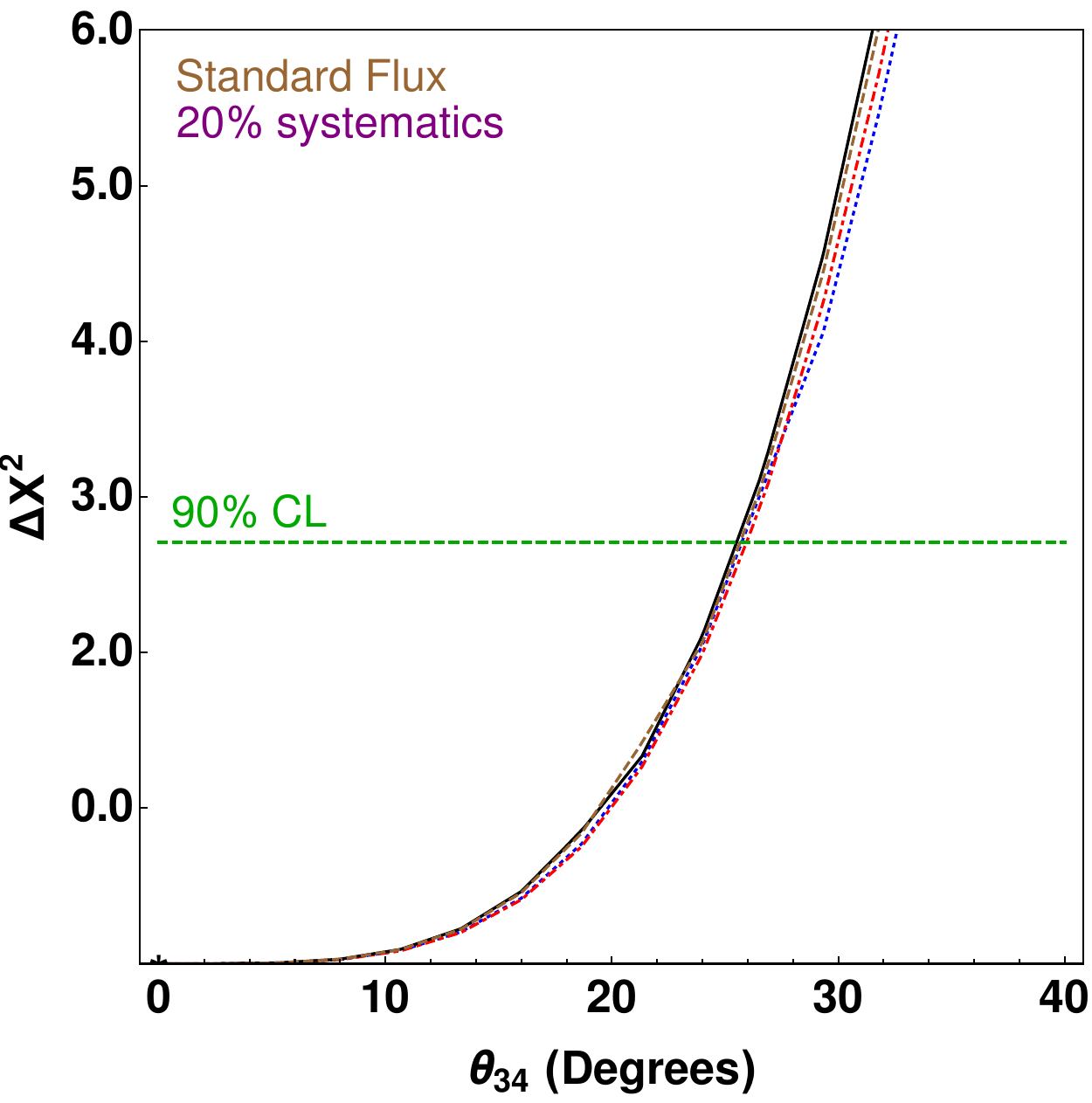} 
\caption{\it  Same as figure \ref{fig:sens34_std} but for the optimized flux.
} 
\label{fig:sens34_opt}
\end{center}
\end{figure}

In figures \ref{fig:sens34_std} (standard flux case) and \ref{fig:sens34_opt} (optimized flux case) we show $\Delta \chi^2$ as a function of  $\theta_{34}$ computed considering all available oscillation channels in DUNE, including the  $\nu_{\mu} \rightarrow \nu_{\tau} $ transition. In the left panels we considered the case where a 10\% of systematic uncertainty is assumed for the signal while the 20\% case is illustrated in the right panel.

As in the previous sections, we present the four different cases corresponding to 6\% of $\tau$ detection efficiency and $S/B = 2.45$ (Red, DotDashed), 18\% efficiency  and again 

\begin{table}[h]
\begin{center}
\begin{tabular}{c|c|c|c|c|c|c|c|c|}
\cline{2-5}
                                                           & \multicolumn{4}{c|}{\textbf{Standard Flux ($10\%$ sys)}}                                                                                                                                                  \\ \cline{2-5} 
                                                           & \multicolumn{2}{c|}{S/B = 2.45}                    & \multicolumn{2}{c|}{S/B = 18.6}                                        \\ \hline
\multicolumn{1}{|c|}{Efficiency of $\nu_{\tau}$ detection}        & $6\%$         & $18\%$        & $6\%$          & $18\%$        \\ \hline
\multicolumn{1}{|c|}{$\theta_{34}$ $^\circ$ ($\Delta m_{41}^2=1 \, eV^2$)} & [0, 26.07]        & [0, 25.52]         & [0, 24.78]        & [0, 24.18] 
\\ \hline
& \multicolumn{4}{c|}{\textbf{Optimized Flux ($10\%$ sys)}}                                                                     \\ \cline{2-5} 
                                                             & \multicolumn{2}{c|}{S/B = 2.45}       & \multicolumn{2}{c|}{S/B = 18.6}                        \\ \hline
\multicolumn{1}{|c|}{Efficiency of $\nu_{\tau}$ detection}        & $6\%$         & $18\%$        & $6\%$          & $18\%$        \\ \hline
\multicolumn{1}{|c|}{$\theta_{34}$ $^\circ$ ($\Delta m_{41}^2=1 \, eV^2$)}  & {[}0, 23.93{]} & {[}0, 22.22{]} & {[}0, 23.47{]} & {[}0, 22.00{]}                              \\ \hline

\end{tabular}
\caption{\label{tab:bounds1} \it  Summary of the bounds at 90\% CL on $\theta_{34}$ (1 degree of freedom) that DUNE may set using 10\% and systematic uncertainties.} 
\end{center}
\end{table}

\begin{table}[h]
\begin{center}
\begin{tabular}{c|c|c|c|c|c|c|c|c|}
\cline{2-5}
                                                           & \multicolumn{4}{c|}{\textbf{Standard Flux ($20\%$ sys)}}                                                                                                                                                   \\ \cline{2-5} 
                                                           & \multicolumn{2}{c|}{S/B = 2.45}                    & \multicolumn{2}{c|}{S/B = 18.6}                                                 \\ \hline
\multicolumn{1}{|c|}{Efficiency of $\nu_{\tau}$ detection}        & $6\%$         & $18\%$        & $6\%$          & $18\%$        \\ \hline
\multicolumn{1}{|c|}{$\theta_{34}$ $^\circ$ ($\Delta m_{41}^2=1 \, eV^2$)} & {[}0, 25.92{]} & {[}0, 25.64{]} & {[}0, 25.72{]} & {[}0, 25.49{]}                                                       \\ \hline
& \multicolumn{4}{c|}{\textbf{Optimized Flux ($20\%$ sys)}}                                                                     \\ \cline{2-5} 
                                                                  & \multicolumn{2}{c|}{S/B = 2.45}       & \multicolumn{2}{c|}{S/B = 18.6}                        \\ \hline
\multicolumn{1}{|c|}{Efficiency of $\nu_{\tau}$ detection}        & $6\%$         & $18\%$        & $6\%$          & $18\%$        \\ \hline
\multicolumn{1}{|c|}{$\theta_{34}$ $^\circ$ ($\Delta m_{41}^2=1 \, eV^2$)} & {[}0, 26.81{]} & {[}0, 25.97{]} & {[}0, 26.64{]} & {[}0, 25.84{]}      \\ \hline

\end{tabular}
\caption{\label{tab:bounds2} \it  Same as table \ref{tab:bounds1} but for a 20\% systematic uncertainty for the $\nu_{\tau}$ appearance channel.} 
\end{center}
\end{table}
$S/B = 2.45$ (Brown, Dashed), 6\% efficiency and $S/B = 18.6$ (Blue, Dotted) and 18\% efficiency and $S/B = 18.6$ (Black, Solid).
In the numerical simulations, the new physics parameters true values have been set to zero except for $\Delta m_{41}^2$ that has been fixed to $1 \, eV^2$. For the standard parameters we considered the central values and uncertainties reported in table \ref{tab:oscpar_nufit2}. The mixing angles $\theta_{14}$ and $\theta_{24}$, together with the two phases $\delta_{1,2}$  have been left free in the fit.

From figures \ref{fig:sens34_std} and \ref{fig:sens34_opt} we extracted the bounds at $90\%$ CL on $\theta_{34}$ that we summarized in tables \ref{tab:bounds1} and  \ref{tab:bounds2}. As expected, the most stringent limit around
22$^\circ$ can be set using the optimized flux with the best efficiency ($18\%$), the best S/B (18.6) and $10\%$ systematic uncertainty, a result in line with the recent NC event analysis carried out in \cite{Coloma:2017ptb}. The importance of $\nu_\tau$ appearance relies on the fact that such a limit would be roughly $20\%$ smaller if only $\nu_{e}$ appearance and $\nu_{\mu}$ disappearance channels 
with standard flux are considered, i.e 27.5$^\circ$.

An important outcome of our analysis is that while the  results with $10\%$ systematic uncertainty and the optimized flux are always better than those obtained with the standard flux, for the $20\%$ systematic this is not always the case; this is because the performances of the other two channels with optimized flux are worse, due to the increased number of $\nu_{\tau}$ CC background events. Thus, even though the $\nu_{\tau}$ appearance can help more in putting strong bounds for the optimized flux case, a larger systematic uncertainty brings to a regime where sensitivity limits are mainly set by the other two channels.

\begin{figure}[h]
\begin{center}
\includegraphics[height=7.5cm,width=7.5cm]{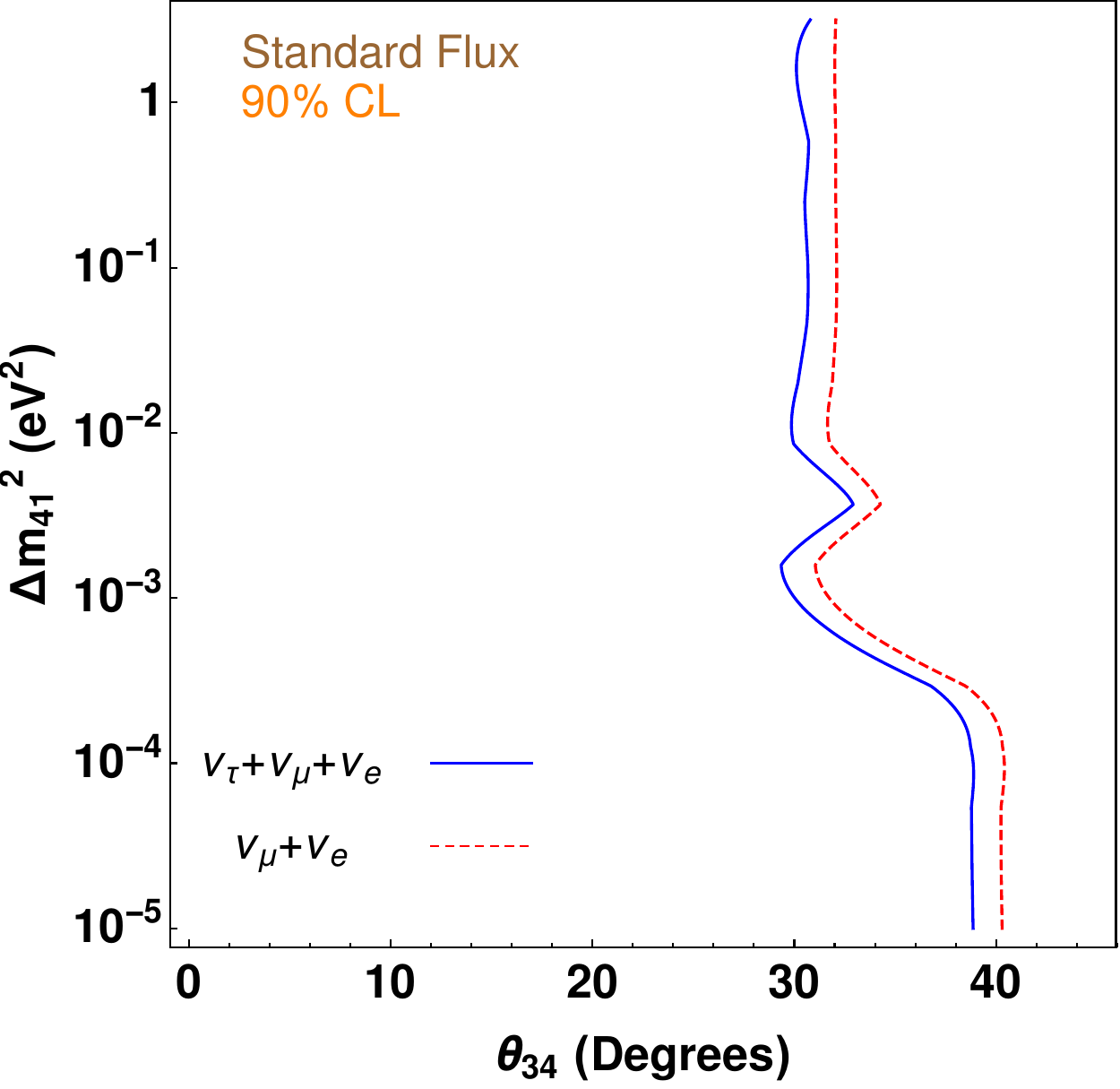} 
\includegraphics[height=7.5cm,width=7.5cm]{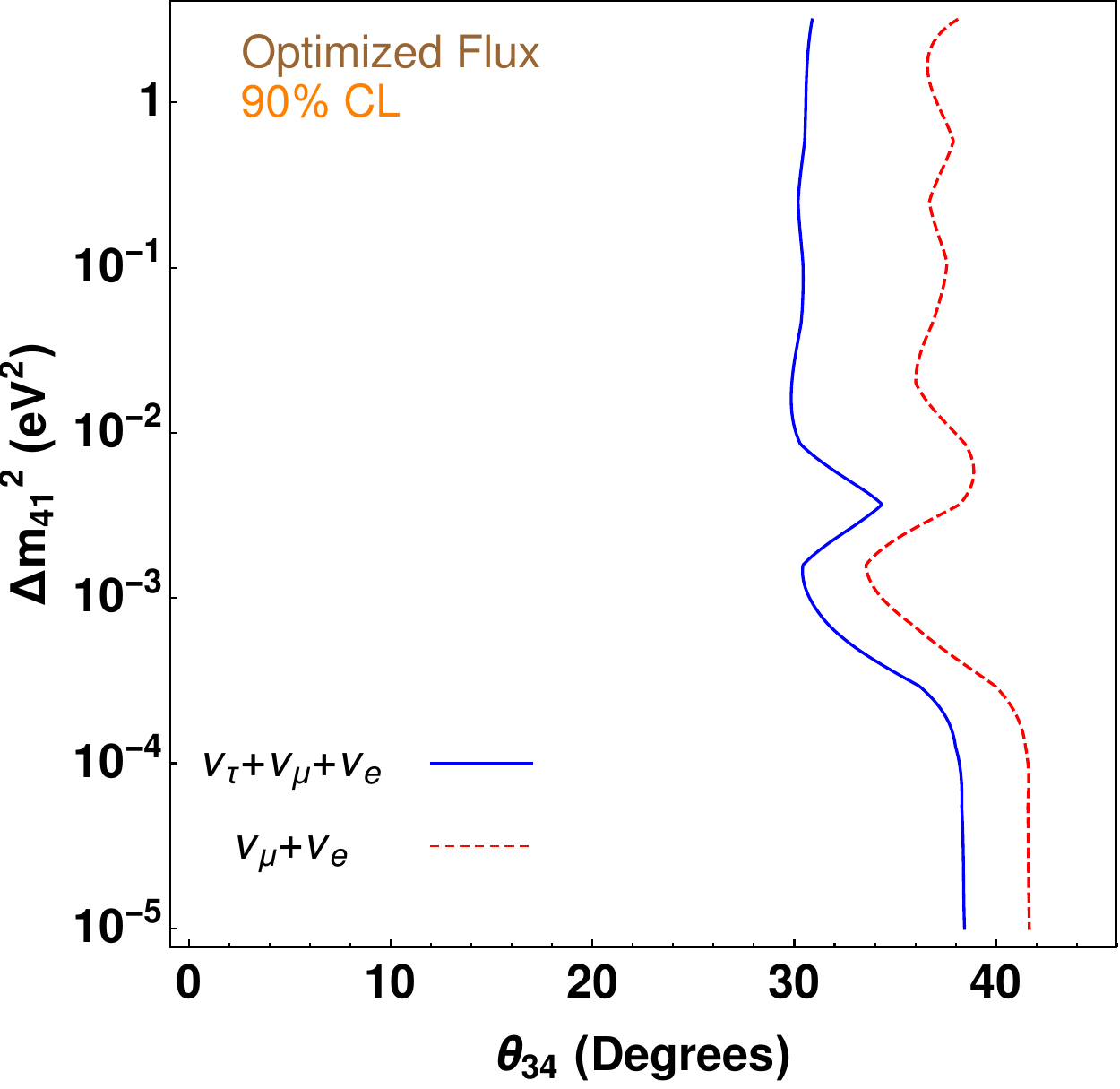} 
\caption{\it  Expected contours at 90\% CL in the $(\theta_{34},\Delta m_{41}^2)$ plane, computed for the standard flux (left panel) and the optimized flux (right panel).  The blue solid lines show the results obtained including the $\nu_{\tau}$ appearance channel while the red dashed lines show the situation with no contribution of the $\tau$ channel. Detection efficiency is 6\% with $S/B = 2.45$ and signal systematic uncertainty of 20\%. All new physics parameters not shown have been left free in the simulation.
} 
\label{fig:corr1_std}
\end{center}
\end{figure}

We complete our anaysis showing the correlation plots at $90\%$ CL in several interesting planes. In all of them, the blue solid lines show the results obtained including the $\nu_{\tau}$ appearance channel in the most conservative case, that is with a 6\% $\tau$ detection efficiency, $S/B = 2.45$ and signal systematic uncertainty of 20\%. The red dashed lines show the situation with no contribution of the $\tau$ channel but only $\nu_{e}$ appearance + $\nu_{\mu}$ disappearance.

In figure \ref{fig:corr1_std} we focused on the $(\theta_{34},\Delta m_{41}^2)$ plane;  we clearly see that for both flux options the gain in sensitivity using the $\tau$ channel is not negligible, amounting to about 6\% and 20\% for the standard and optimized flux cases, respectively.

\begin{figure}[h]
\begin{center}
\includegraphics[height=7.5cm,width=7.5cm]{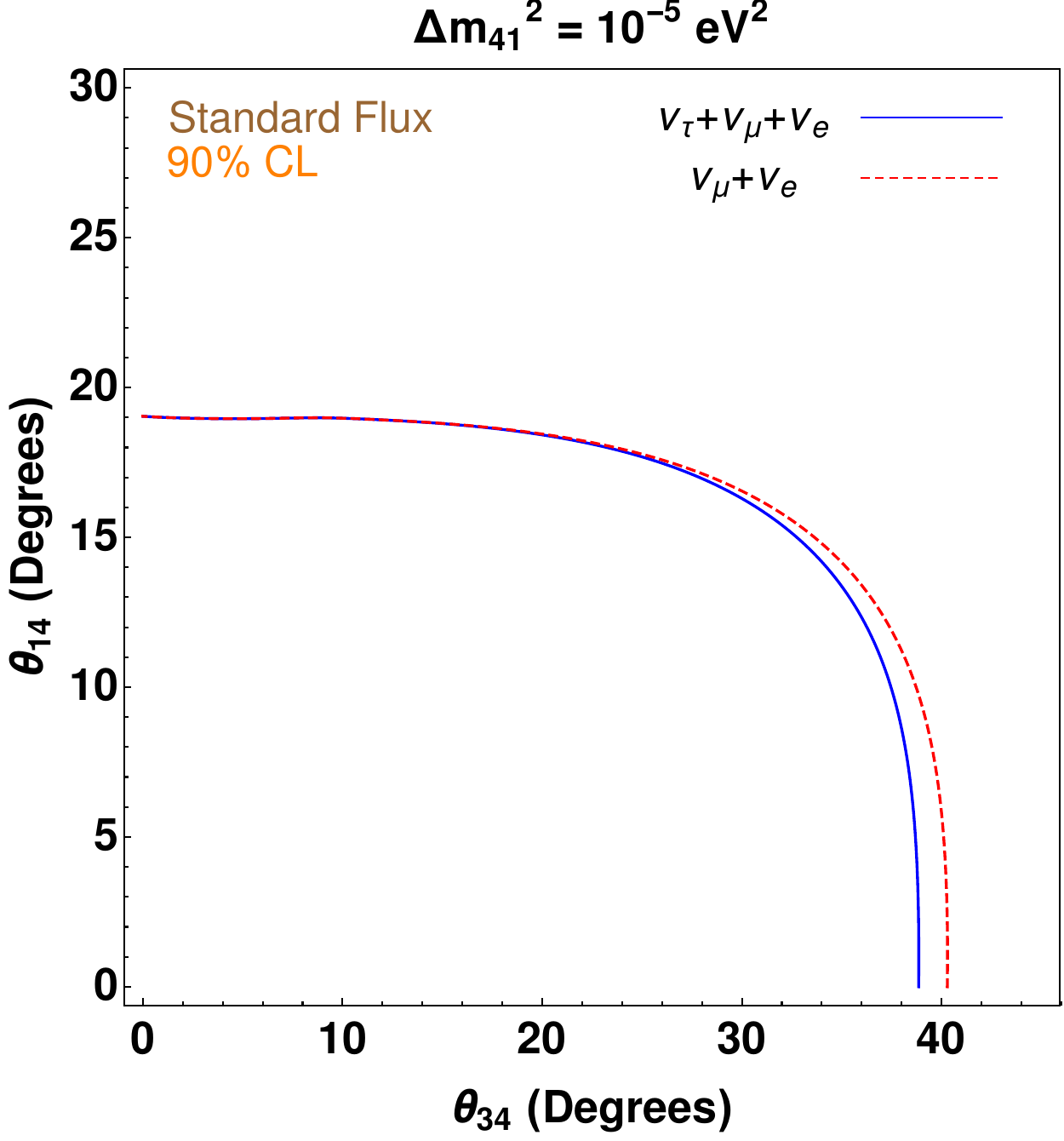} 
\includegraphics[height=7.5cm,width=7.5cm]{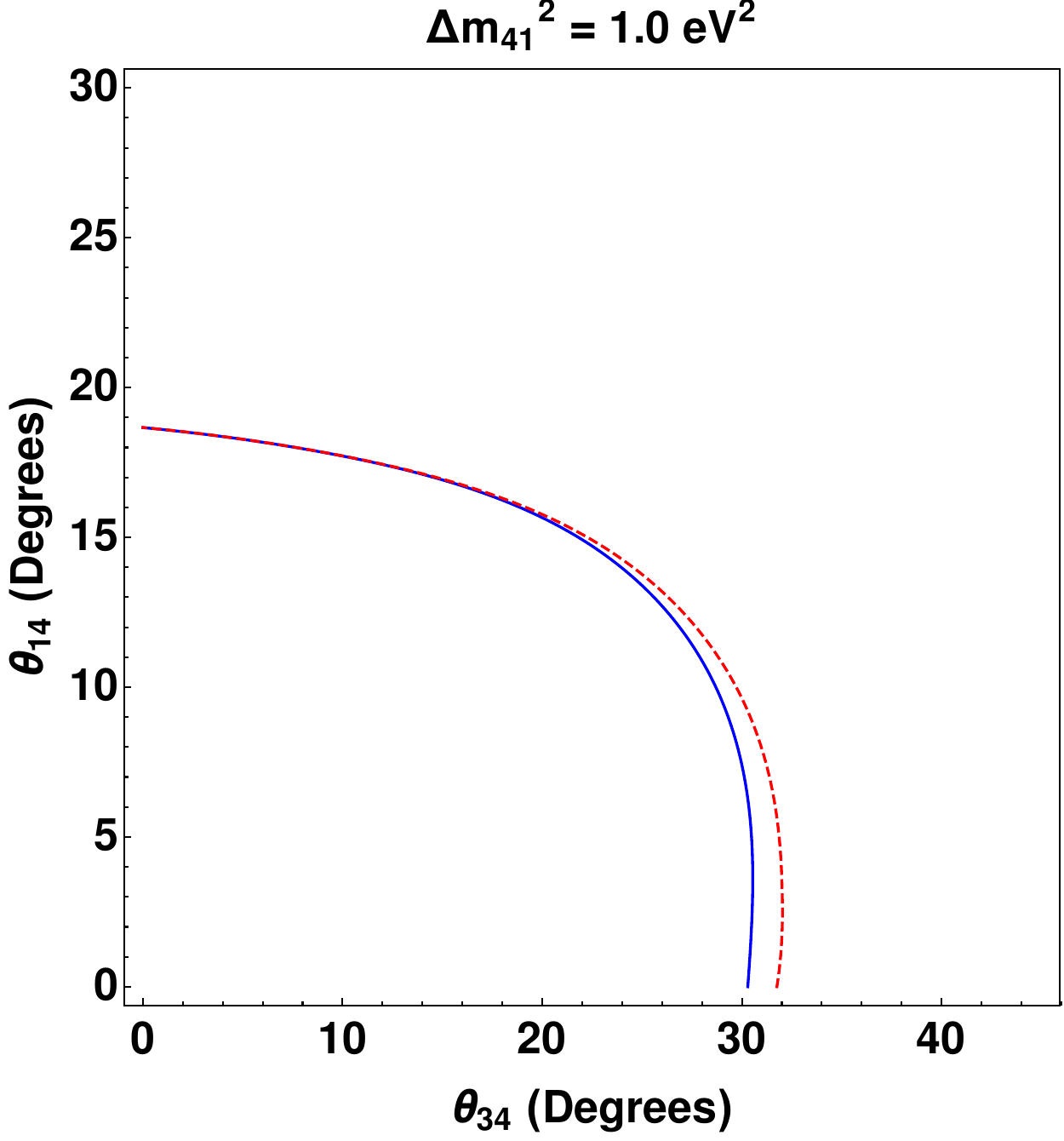} 
\includegraphics[height=7.5cm,width=7.5cm]{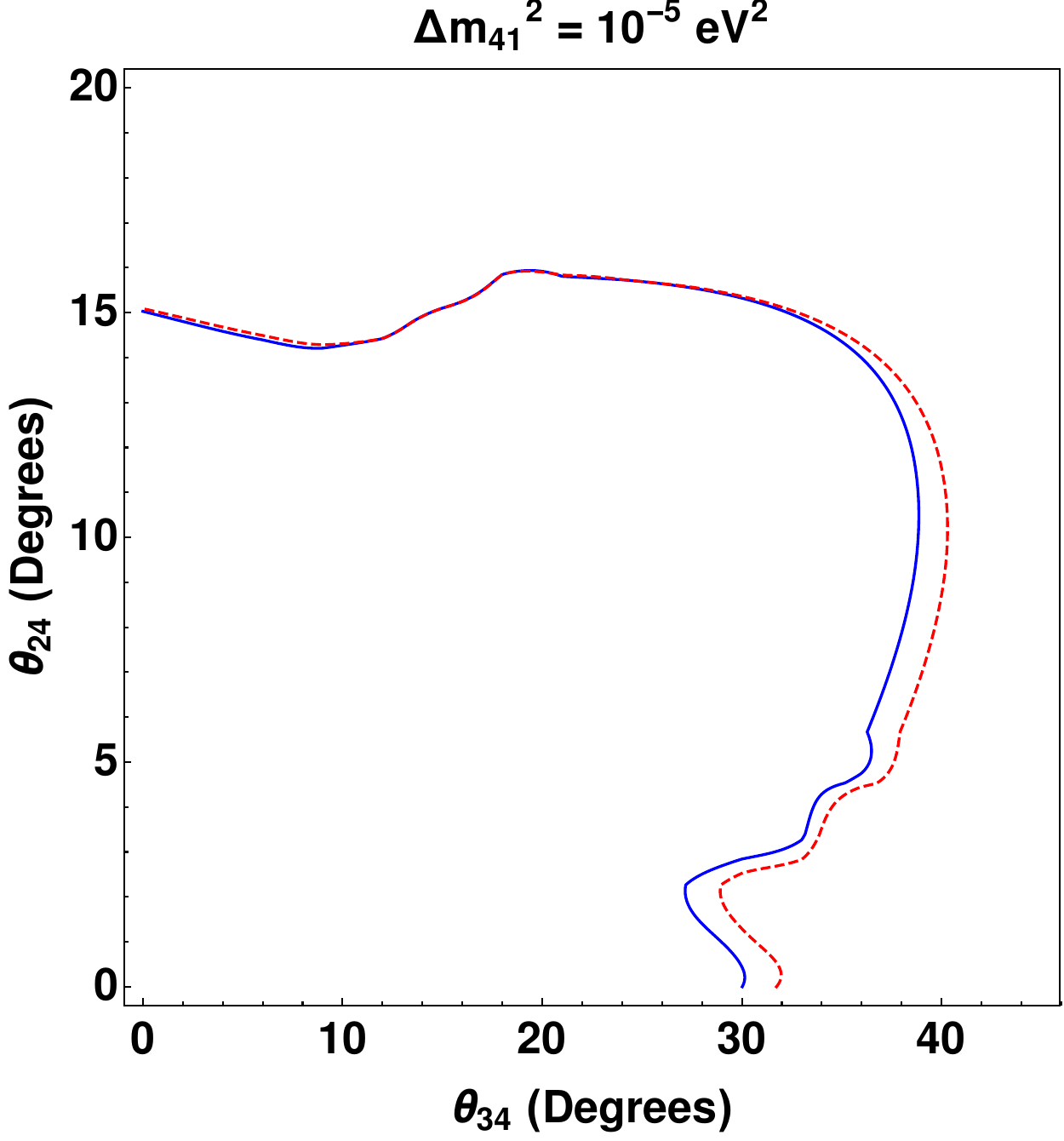} 
\includegraphics[height=7.5cm,width=7.5cm]{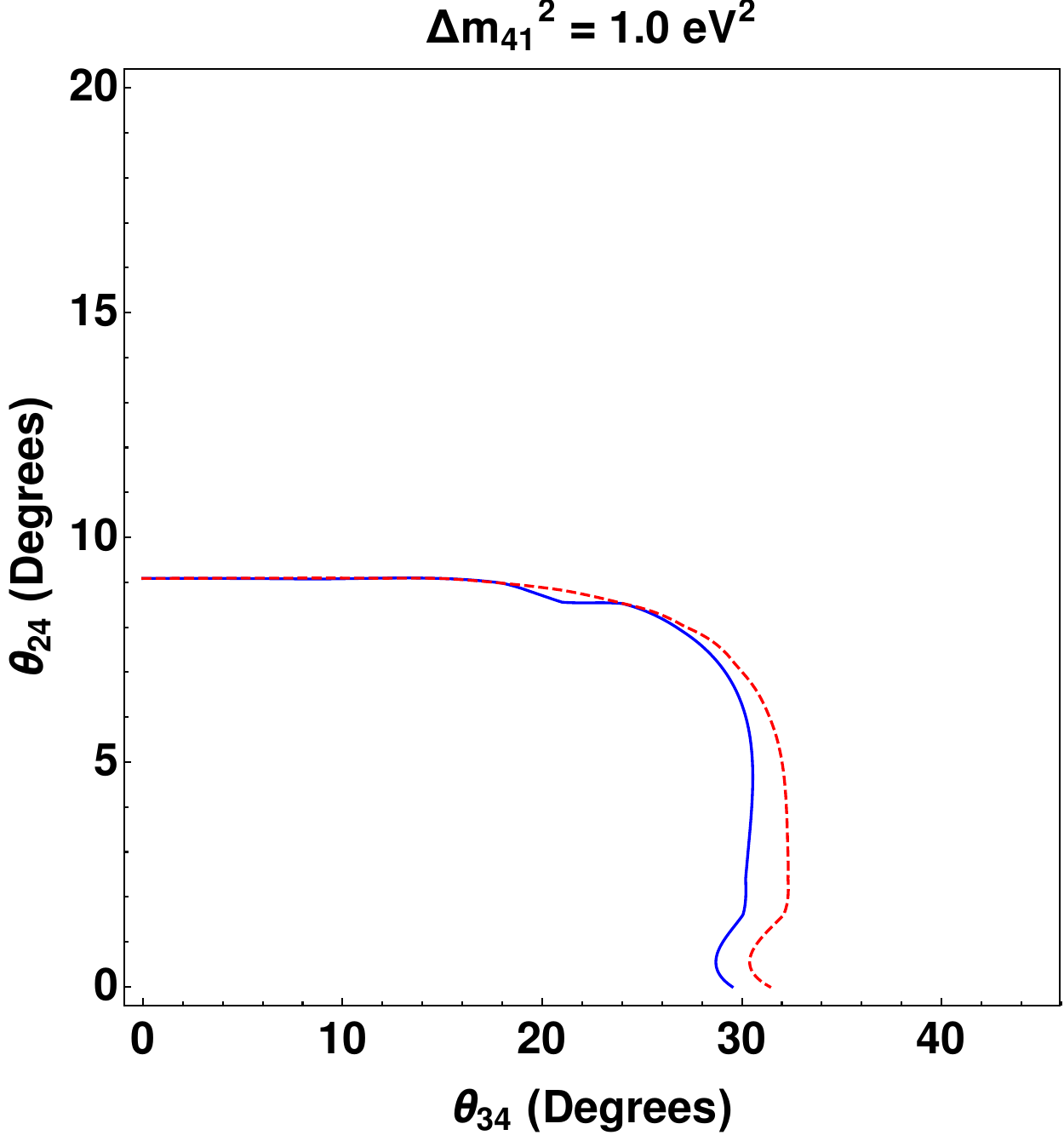} 
\caption{\it  Same as figure \ref{fig:corr1_std} but in the planes $(\theta_{34},\theta_{14})$ (upper panels) and $(\theta_{34},\theta_{24})$ (lower panels). The new mass difference $\Delta m_{41}^2$ is set to two distinct values: $10^{-5} \, eV^2$ (left panels) and $1  \, eV^2$ (right panels). Standard flux is assumed.} 
\label{fig:corr2_std}
\end{center}
\end{figure}

\begin{figure}[h]
\begin{center}
\includegraphics[height=7.5cm,width=7.5cm]{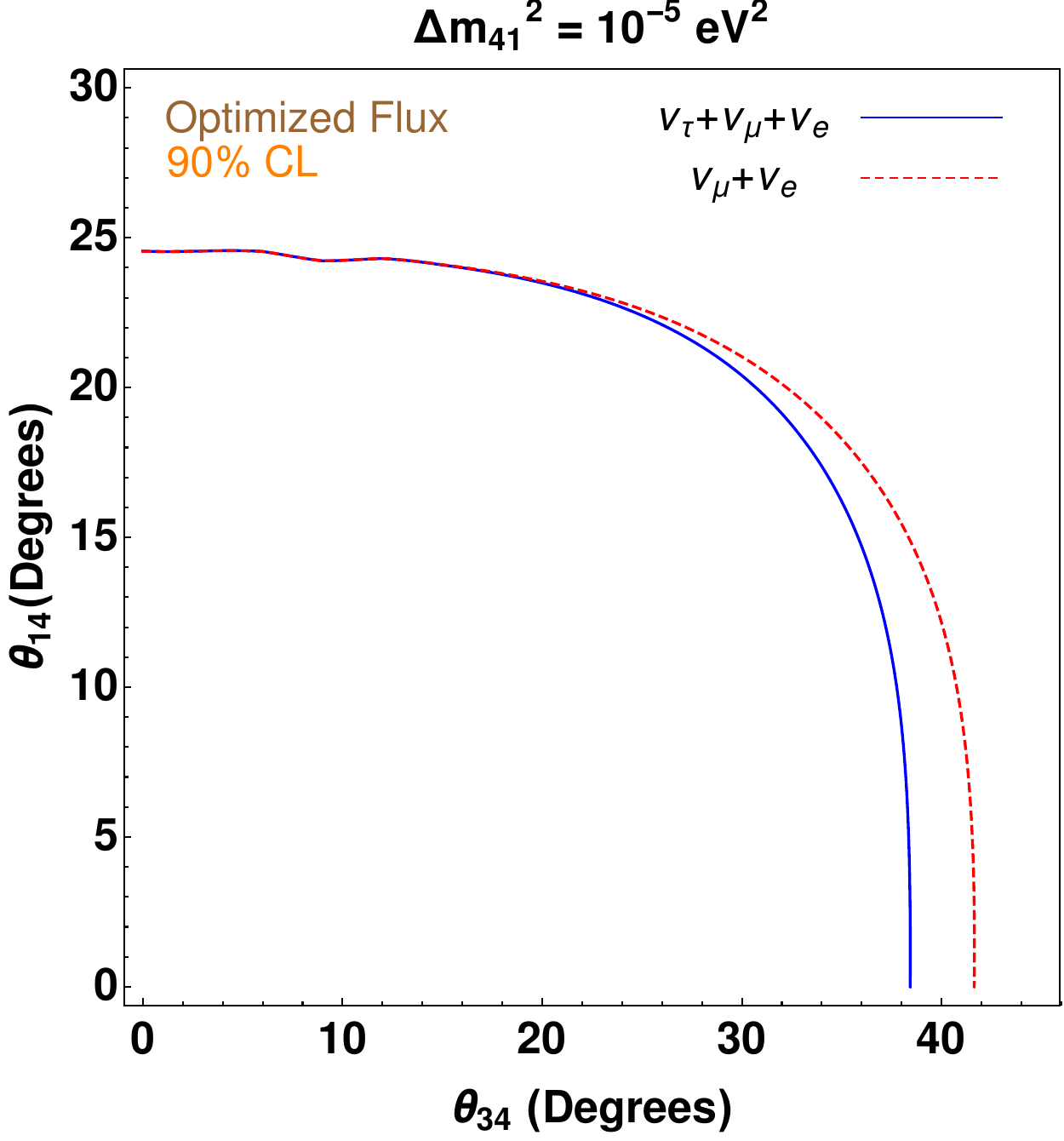} 
\includegraphics[height=7.5cm,width=7.5cm]{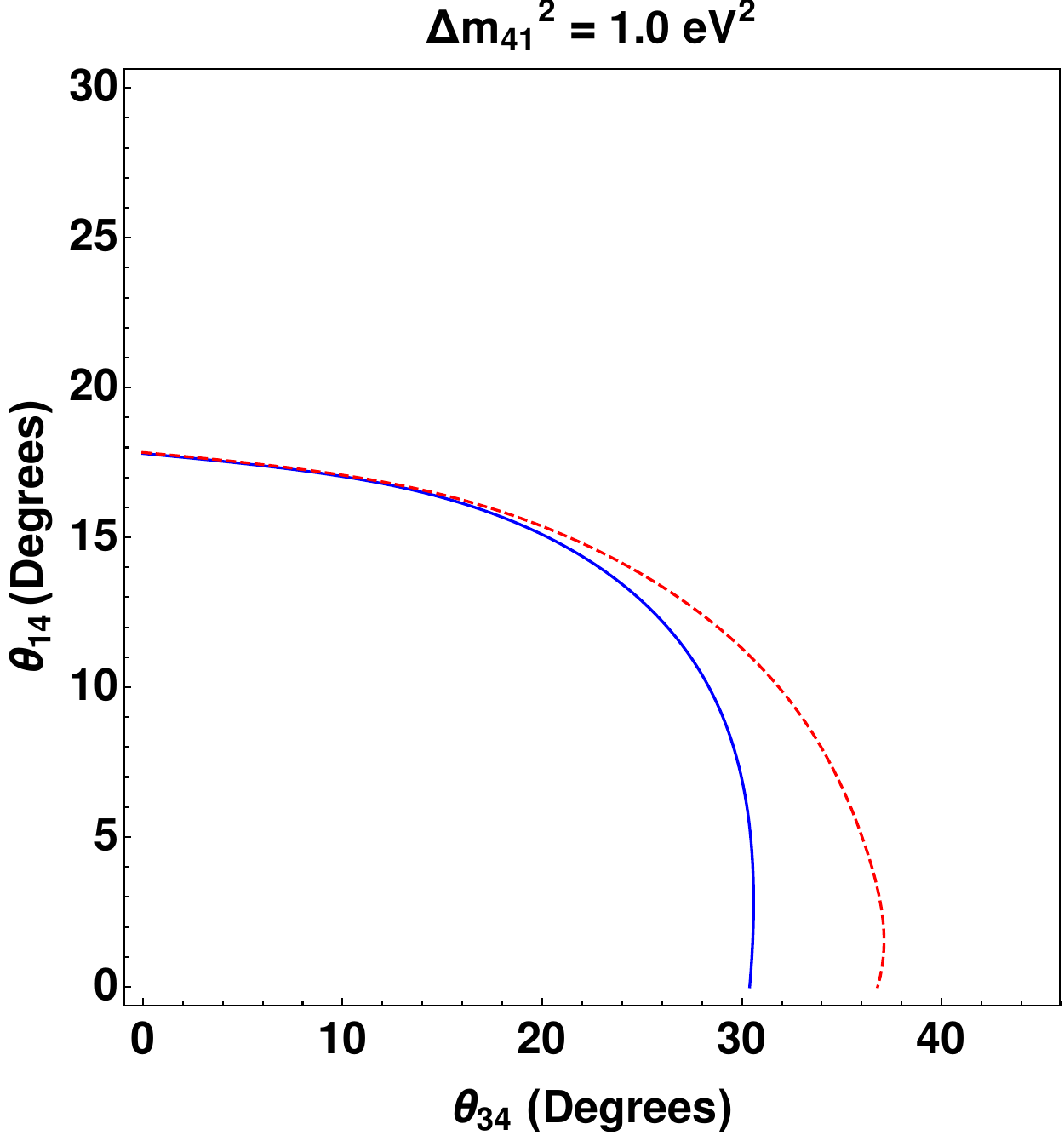} 
\includegraphics[height=7.5cm,width=7.5cm]{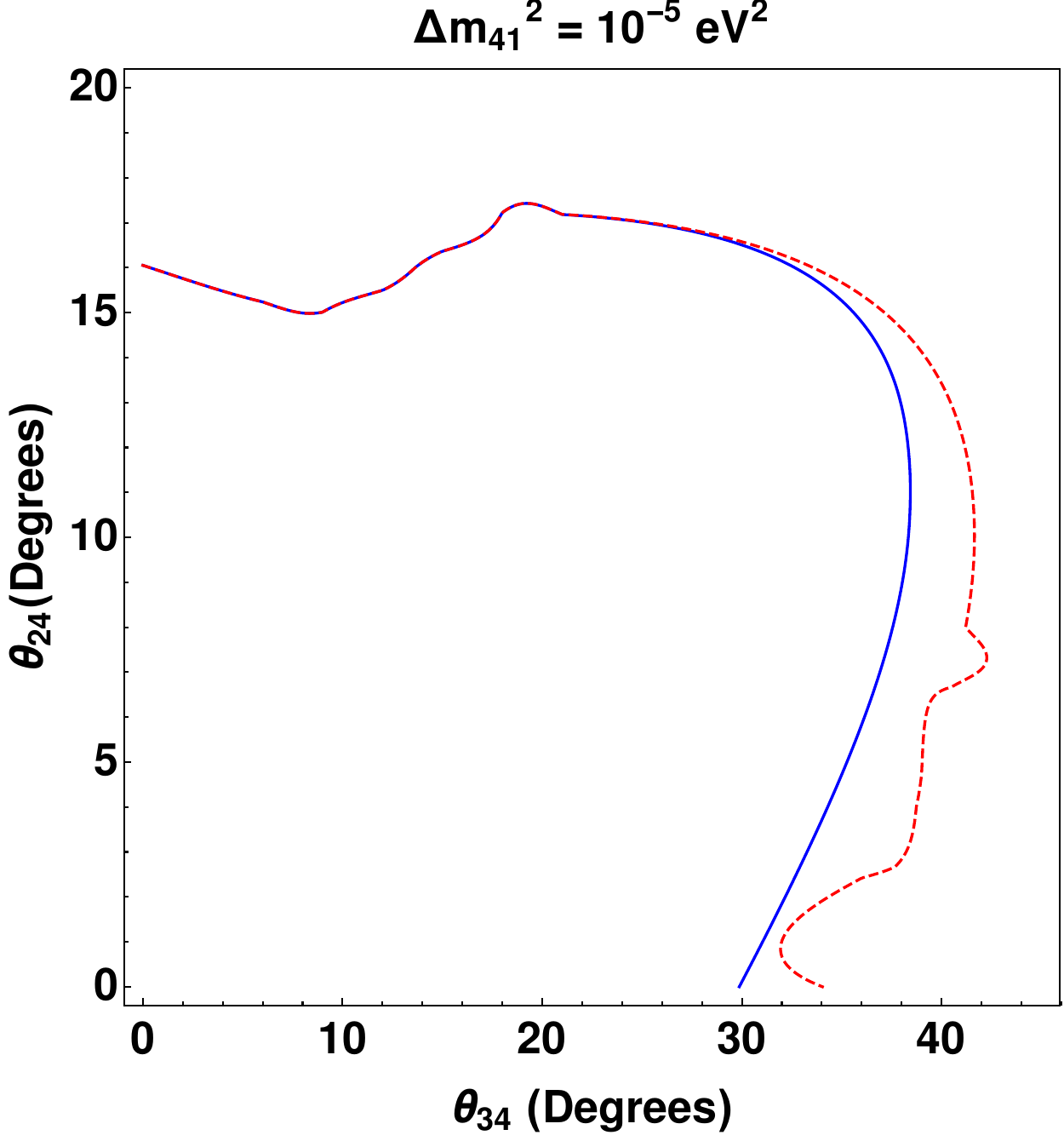} 
\includegraphics[height=7.5cm,width=7.5cm]{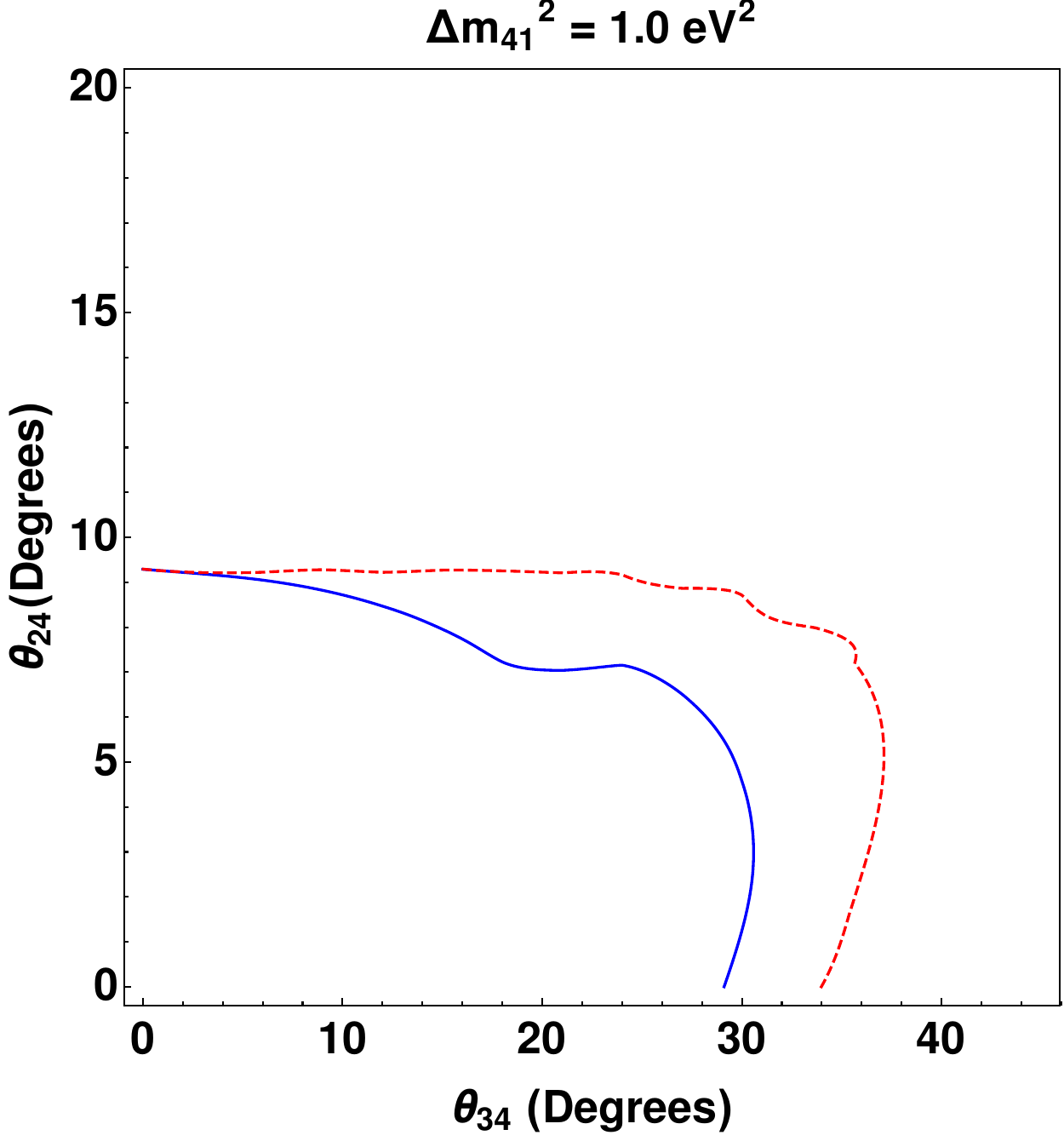} 
\caption{\it   Same as figure \ref{fig:corr2_std} but for the optimized flux.
} 
\label{fig:corr2_opt}
\end{center}
\end{figure} 
In figure \ref{fig:corr2_std} we show the correlation plots in the planes $(\theta_{34},\theta_{14})$ (upper panels) and $(\theta_{34},\theta_{24})$ (lower panels). The new mass difference $\Delta m_{41}^2$ is set to $10^{-5} \, eV^2$ (left panels) and $1  \, eV^2$ (right panels) and the standard flux is assumed. First of all, we confirm that the $\tau$ channel does not significantly improve the bounds on $\theta_{14}$ and $\theta_{24}$, which are around 20$^\circ$ and between 10$^\circ$ and 20$^\circ$, respectively. Instead, the upper limit on $\theta_{34}$ can be improved by roughly 7\% for both values of $\Delta m_{41}^2$ analyzed here. As for the absolute upper bound on $\theta_{34}$, we observe that  when the mass splitting is smaller, $\theta_{34}$ is confined to larger values  because, as seen from the expression in 
eq.(\ref{mu-taulight}), in this mass regime the non standard matrix elements are always sub leading. 

The situation where the optimized flux is taken into account is presented in 
figure \ref{fig:corr2_opt}; consistently with figure \ref{fig:corr1_std}, the final sensitivity to $\theta_{34}$ when adding to the simulations all three transition channels is very similar to that obtained for the standard case. The real improvement carried by the optimized flux are on the (bad) limits set by $\nu_{e}$ appearance + $\nu_{\mu}$ disappearance only, since they suffer by a large $\tau$ CC background.
Finally, we comment on the fact that the sensitivities to the other sterile neutrino parameters are not affected at all by the addition of the $\nu_\tau$ appearance channel.

\section{Discussion and Conclusions}
\label{concl}

The DUNE experiment is being proposed as a high precision next-generation neutrino experiment to be built in the USA. The baseline of DUNE is suitable for observing the neutrino mass
hierarchy and measuring the CP phase $\delta_{CP}$ but only $\nu_e$ appearance and $\nu_{\mu}$ disappearance have been considered in most of the analysis.
We studied the impact of the inclusion of $\nu_{\mu} \rightarrow \nu_{\tau}$ and $\bar{\nu}_\mu \rightarrow \bar{\nu}_{\tau}$ channels in the DUNE experiment, 
considering the electronic $\tau$ decay mode.
This oscillation channel is interesting because, on the one hand, it can confirm the three neutrino paradigm thanks to the redundancy produced by unitarity of the $U_{PMNS}$ and, on the other hand, it can probe several scenarios of new physics thanks to the almost unique dependence of the transition probability on certain non-standard parameters.

Our simulations have taken into account the impact of various systematic uncertainties, $\nu_{\tau}$ detection efficiencies, two different $S/B$ ratios and two distinct neutrino fluxes on the sensitivities of the oscillation parameters.
(3.5 + 3.5) years of data taking for both flux options have been considered. Based on  the simulations performed with the GLoBES software presented in sections \ref{stphys}, \ref{sect:nsi} and \ref{sec:sterile}, we conclude the following:

\begin{itemize}
 \item For the standard physics, the addition of $\nu_{\tau}$ appearance channel       
       does not improve the sensitivities of any of the neutrino oscillation parameter 
       set by the other two channels already being considered in DUNE.
\item       The performances of the tau optimized flux in the $\nu_e$ appearance and $\nu_{\mu}$ disappearance channels
       result generally in a worsening of the sensitivities, thus  overshadowing the 
       advantage one may get from the increase in the $\nu_{\tau}$ statistics. This is mainly due to the increased background events in both the $\nu_e$ and $\nu_{\mu}$ 
       channels. 
 \item In the new physics cases, the NSI parameter sensitivities are little  affected by the addition of the new channel, except for the coupling $|\epsilon_{\mu\tau}|$ 
       for which improved limits (about 15\% better than the no-$\tau$ case) can be found.
 \item For the sterile neutrino 3+1 case, the only parameter that shows an increase in sensitivity is the mixing angle $\theta_{34}$ whose 
       improvement is about 20\% compared to the case where $\tau$ signal events are not considered.
\end{itemize}

We remark that it may be possible to improve the above-mentioned sensitivities if we add to this analysis the tau hadronic decay channel as well \cite{deGouvea:2019ozk}, which has a  better $\nu_{\tau}$ detection efficiency
and consequently a bigger statistics.

\acknowledgments

A.G. is grateful to Alessio Barbensi for various support; D. M. thanks Enrique Fernandez-Martinez and Pedro Pasquini for useful discussions.

\end{document}